\documentclass[a4paper, 12pt, version-1-compatibility]{article}
\usepackage[margin=2cm]{geometry}
\usepackage[utf8]{inputenc}
\usepackage[dvipsnames]{xcolor}
\usepackage{hyperref}
\usepackage{amsmath}
\usepackage{amssymb}
\usepackage{booktabs}
\usepackage{xcolor}
\usepackage{footnote}
\makesavenoteenv{tabular}
\makesavenoteenv{table}
\usepackage{graphicx}
\graphicspath{{img/}}
\usepackage{subcaption}
\usepackage{textcomp}
\usepackage{gensymb}
\usepackage{makecell}
\usepackage[ruled]{algorithm2e}
\usepackage{soul}
\usepackage{tikz}
\usetikzlibrary{fit}
\usetikzlibrary{backgrounds}
\tikzset{myarrow/.style = {-{Triangle[width=5,angle'=45]}, line width=0.75}}
\usepackage{tikz}
\usetikzlibrary{shapes,arrows,decorations,backgrounds}
\usetikzlibrary{mindmap,trees}
\usetikzlibrary{decorations.pathreplacing}
\usetikzlibrary{plotmarks}
\usetikzlibrary{calc}
\usetikzlibrary{fadings}
\usepackage{pgflibraryshapes}  
\usetikzlibrary{arrows}
\usepackage[version-1-compatibility]{siunitx}
\tikzstyle{pblock} = [rectangle, draw, fill=yellow!20, text width=5.5em, text centered, minimum height=5em, node distance=7.5cm]
\tikzstyle{void} = [fill=white, node distance=5cm]
\tikzstyle{connector} = [->,thick]
\tikzset{cross/.style={cross out, draw, 
		minimum size=2*(#1-\pgflinewidth), 
		inner sep=0pt, outer sep=0pt}}

\usetikzlibrary{shapes.geometric}
\usetikzlibrary{decorations.pathmorphing}

\usepackage{verbatim}
\usepackage[siunitx]{circuitikz}
\usetikzlibrary{positioning}

\definecolor{outputcolor}{HTML}{5C3046}

\definecolor{matplotlibblue}{HTML}{1f77b4}
\definecolor{matplotliborange}{HTML}{ff7f0e}

\definecolor{inputcolor}{HTML}{228B22}
\definecolor{outputcolor}{HTML}{045174}
\definecolor{hiddencolor}{HTML}{E87A00}

\def\gunwidth{0.2927cm}
\def\Lone{3.0cm}
\def\LSone{4.0cm}
\def\Ltwo{5.31cm}
\def\LStwo{6.15cm}
\def\Lthree{6.9cm}
\def\LSthree{8.2cm}
\def\Lfive{9.5cm}

\def\cavitylength{1.20713cm}
\def\lslength{1.0}

\def\cavityheight{0.2cm}
\def\solenoidheight{0.3cm}
\def\solenoidshift{0.4cm}

\newcommand{\cavityhalf}{
    \def\bumpwidth{ {\cavitylength / (1.5 * 7)} }
    \def\sep{\bumpwidth / 2}

    \draw (0, 0) -- +(0, \cavityheight) \foreach \i in {0, ..., 5} {  arc [start angle=180, end angle=0, radius=\bumpwidth / 2] -- +(\sep, 0)} arc [start angle=180, end angle=0, radius=\bumpwidth / 2] -- +(0, -\cavityheight);
}

\newcommand{\cavity}[3]{
    \begin{scope}[xshift=#1, yshift=#2, color=matplotliborange, thick]
        \cavityhalf
        \begin{scope}[yscale=-1]
            \cavityhalf
        \end{scope}
        \node[yshift=-0.12cm, xshift=-0.025cm, align=center, black] at (\cavitylength / 2, \cavityheight / 2) {\small #3};
    \end{scope}
}

\newcommand{\solenoidbox}{
    \begin{scope}[thick, draw=matplotlibblue, fill=matplotlibblue!20]
        \filldraw (0, 0) rectangle +(1cm, \solenoidheight);
    \end{scope}
}
\newcommand{\solenoid}[1]{
    \begin{scope}[xshift=#1]
        \begin{scope}[yshift=\solenoidshift]
            \solenoidbox
        \end{scope}
        \begin{scope}[yshift=-\solenoidshift - \solenoidheight]
            \solenoidbox
        \end{scope}
    \end{scope}
}

\usepackage[style=numeric, sorting=none]{biblatex}
\def\ga{genetic algorithm}
\def\gas{genetic algorithms}
\def\forwardmodel{forward model}
\def\invertiblemodel{invertible model}
\def\isodar{IsoDAR}
\def\opal{OPAL}
\def\dvar{design variable}
\def\dvars{design variables}
\def\qois{quantities of interest}

\newcommand{\figref}[1]{Fig.\xspace\ref{#1}}
\newcommand{\tabref}[1]{Table~\ref{#1}}
\newcommand{\eqnref}[1]{Eq.~\ref{#1}}
\newcommand{\secref}[1]{Section~\ref{#1}}

\title{Fast,  efficient  and  flexible  particle  accelerator  optimisation using densely connected and invertible neural networks}
\author{Renato Bellotti, Romana Boiger \& Andreas Adelmann}
\date{Paul Scherrer Institut, 5232 Villigen, Switzerland}

\newcommand{\myvec}[1]{\mathbf{#1}}

\newcommand{\htp}{\ensuremath{\mathrm{H}_2^+}\xspace}

\definecolor{matplotlibblue}{HTML}{1f77b4}
\definecolor{matplotliborange}{HTML}{ff7f0e}

\DeclareMathOperator{\percentile}{\mathrm{percentile}}

\usepackage{lmodern}

\begin{document}
\maketitle
\begin{abstract}
\noindent Particle accelerators are enabling tools for scientific exploration and discovery in various disciplines. 
Finding optimized operation points for these complex machines is a challenging task, however, due to the large number of parameters involved and the underlying non-linear dynamics. Here, we introduce two families of data-driven surrogate models, based on deep and invertible neural networks, that can replace the expensive physics computer models. 

These models are employed in multi-objective optimisations to find Pareto optimal operation points for two fundamentally different types of particle accelerators. Our approach reduces the time-to-solution for a multi-objective accelerator optimisation up to a factor of $640$ and the computational cost up to $98\%$.\ The framework established here should pave the way for future on-line and real-time multi-objective optimisation of particle accelerators.
\end{abstract}

\section*{Introduction}
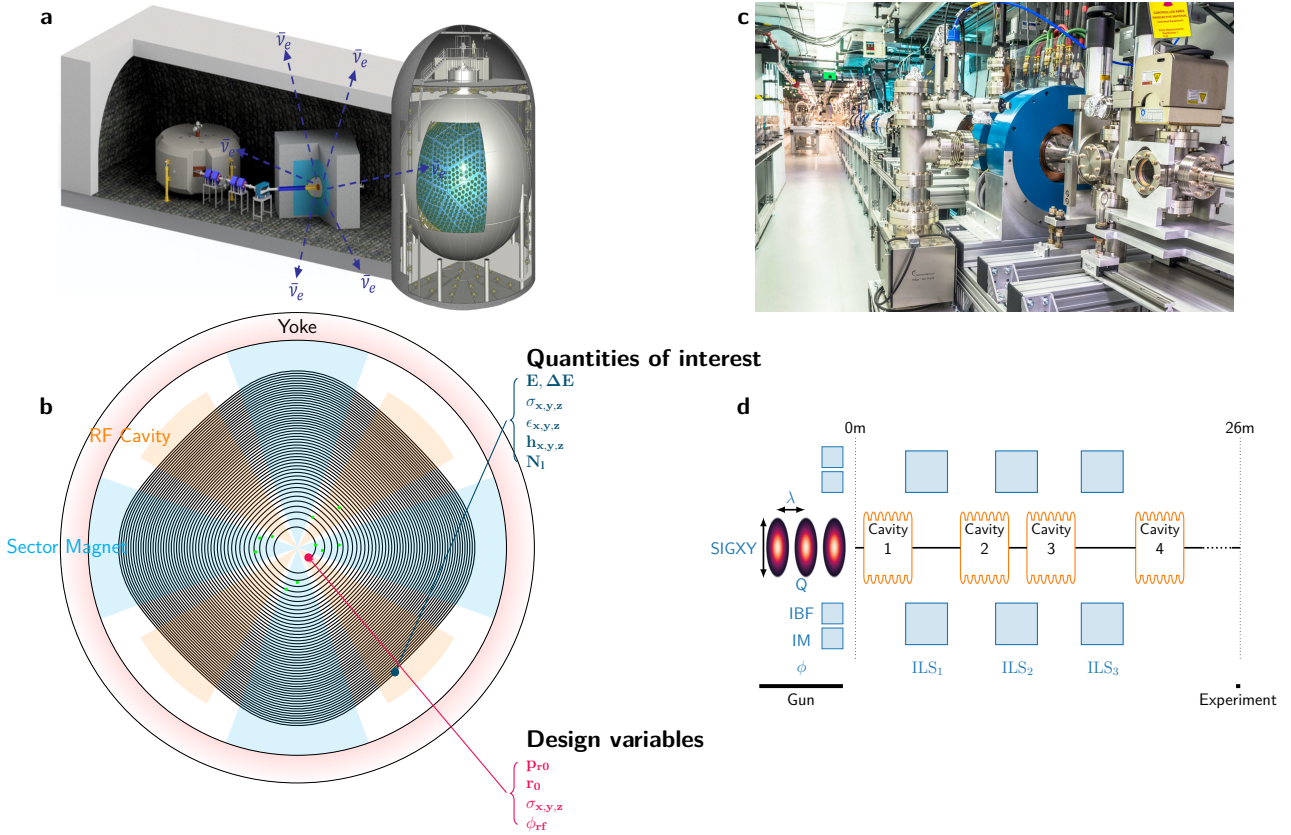
\begin{figure}
\resizebox{\textwidth}{!}{
\begin{tikzpicture}[scale=3.5]
%
%
\begin{scope}[xscale=1,yscale=1,xshift=0,yshift=5] 
\draw[thick,rotate=90] plot file {newData.dat} ; 

\fill[cyan, opacity=0.15] (0, 0) -- ({1.5*cos(110)}, {1.5*sin(110)}) arc (110:70:1.5);
\fill[cyan, opacity=0.15] (0, 0) -- ({1.5*cos(20)}, {1.5*sin(20)}) arc (20:-20:1.5);
\fill[cyan, opacity=0.15] (0, 0) -- ({1.5*cos(-110)}, {1.5*sin(-70)}) arc (-110:-70:1.5);
\fill[cyan, opacity=0.15] (0, 0) -- ({-1.5*cos(20)}, {-1.5*sin(-20)}) arc (160:200:1.5);
\fill[orange, opacity=0.15] (0, 0) -- ({1.3*cos(60)}, {1.3*sin(60)}) arc (60:30:1.3);
\fill[orange, opacity=0.15] (0, 0) -- ({1.3*cos(-30)}, {1.3*sin(-30)}) arc (-30:-60:1.3);
\fill[orange, opacity=0.15] (0, 0) -- ({1.3*cos(210)}, {1.3*sin(210)}) arc (210:240:1.3);
\fill[orange, opacity=0.15] (0, 0) -- ({1.3*cos(150)}, {1.3*sin(150)}) arc (150:120:1.3);

\filldraw[even odd rule,inner color=red,outer color=white] 
(0,0) circle (1.5)
(0,0) circle (1.7);
\node at (0, 1.6) {\large Yoke};

\node [orange] (rf) at (-1.2, 0.8) {\large RF Cavity};

\node [cyan] (hill) at (-1.65, 0) {\large Sector Magnet};

\fill[OrangeRed] (0.08, -0.07) circle (0.03);
\fill[outputcolor] (0.7, -0.9) circle (0.03);

\node[align=left, anchor=west] (output) at (1.6, 1) {\Large \textbf{Quantities of interest}\\
\textcolor{outputcolor}{$\mathbf{E, \Delta E}$}\\ 
\textcolor{outputcolor}{$\mathbf{\sigma_{x, y, z}}$}\\ 
\textcolor{outputcolor}{$\mathbf{\epsilon_{x, y, z}}$}\\ 
\textcolor{outputcolor}{$\mathbf{h_{x, y, z}}$}\\
\textcolor{outputcolor}{$\mathbf{N_l}$}};

\draw [decorate,decoration={brace,amplitude=5pt},xshift=-4pt,outputcolor,thick]
(1.72, 0.62) -- (1.72,1.225) node [OrangeRed,midway] {};

\draw [->, >=stealth, outputcolor, thick] (1.5, 0.92) -- (0.7, -0.9); 

\node[align=left, below=9.4cm of output.west, anchor=west] 
{\Large \textbf{Design variables}\\ 
\textcolor{OrangeRed}{$\mathbf{p_{r0}}$}\\ 
\textcolor{OrangeRed}{$\mathbf{r_0}$}\\
\textcolor{OrangeRed}{$\mathbf{\sigma_{x, y, z}}$}\\
\textcolor{OrangeRed}{$\mathbf{\phi_{rf}}$}};

\draw [decorate,decoration={brace,amplitude=5pt},xshift=-4pt,OrangeRed,thick]
(1.72,-2) -- (1.72,-1.55) node [OrangeRed,midway] {};

\draw [->, >=stealth, OrangeRed, thick] (1.5, -1.77) -- (0.08, -0.07); 

\foreach \position in {(0, -0.25),(-0.08, -0.3), (-0.3, -0.03), (-0.27, 0.07), (-0.18, 0.08), (0.3, 0.29), (0.4, 3.1), (0.11, 0.22), (0.13, 0.02), (0.18, -0.02), (0.3, 0.02)}
{
	\draw \position node[cross=2pt, very thick, green] {};
}
\end{scope}

\begin{scope}[xscale=0.3,yscale=1,xshift=300,yshift=5] 
\node[inner sep=0pt, xshift=1.7cm] (gaussians) {\includegraphics[width=2cm]{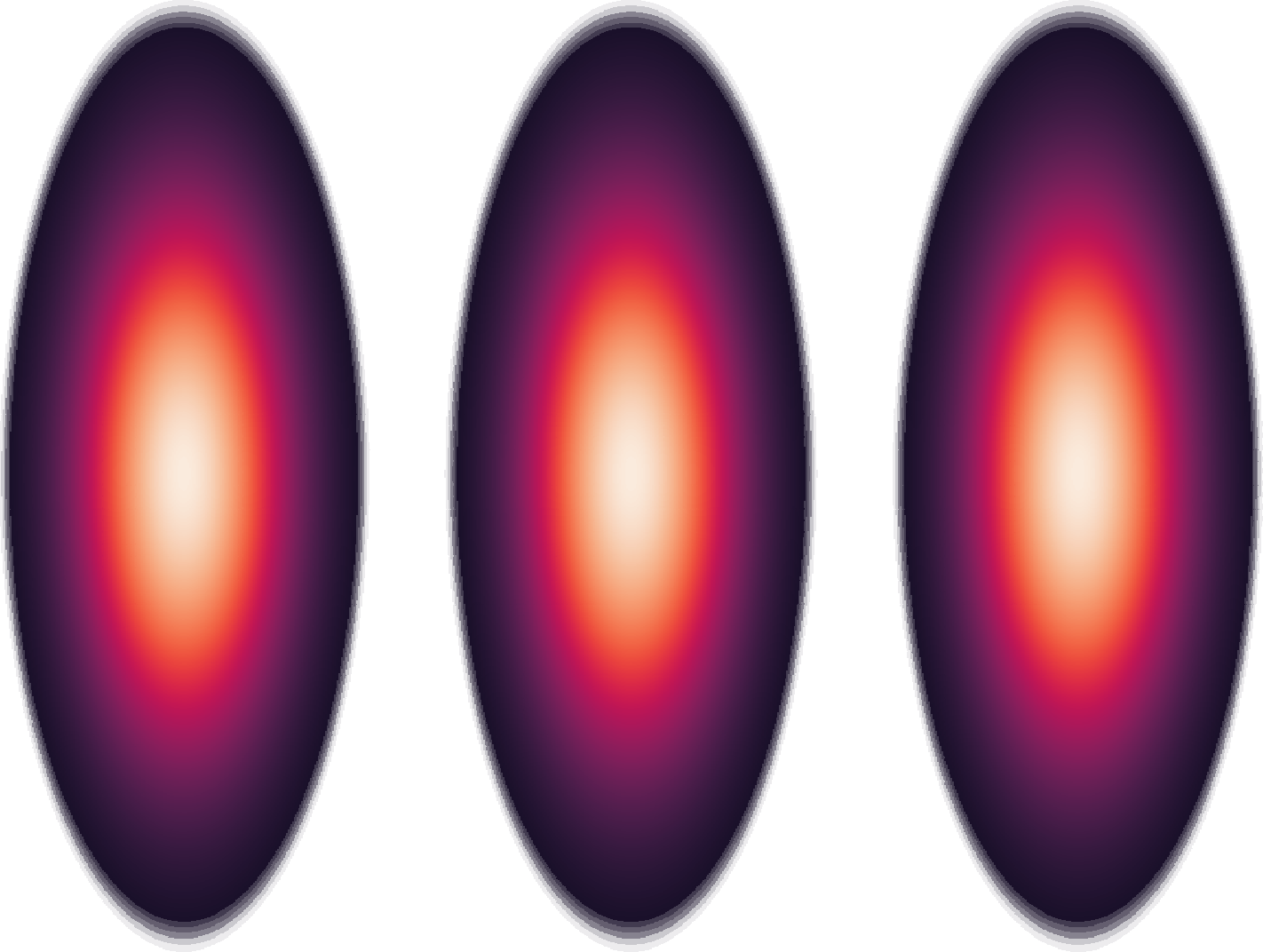}};
\draw[very thick, latex-latex] (0.9cm, 0.275cm) -- +(0.7cm, 0cm) node[color=matplotlibblue,anchor=south, pos=0.5]{$\mathbf{\lambda}$};
\draw[very thick, latex-latex] (0.6cm, 0.22cm) -- +(0cm, -0.44cm) node[color=matplotlibblue,anchor=east, pos=0.5] {SIGXY};

\filldraw[draw=matplotlibblue, fill=matplotlibblue!20] (2cm, \solenoidshift) rectangle +(0.5cm, 0.5*\solenoidheight);
\filldraw[draw=matplotlibblue, fill=matplotlibblue!20] (2cm, -\solenoidshift - 0.5*\solenoidheight) rectangle +(0.5cm, 0.5*\solenoidheight);

\filldraw[draw=matplotlibblue, fill=matplotlibblue!20] (2cm, \solenoidshift + 0.6*\solenoidheight) rectangle +(0.5cm, 0.5*\solenoidheight);
\filldraw[draw=matplotlibblue, fill=matplotlibblue!20] (2cm, -\solenoidshift - 0.5*\solenoidheight - 0.6*\solenoidheight) rectangle +(0.5cm, 0.5*\solenoidheight);
\node[color=matplotlibblue, xshift=-0.3cm] at (1.8cm, -0.27cm) {Q};
\node[color=matplotlibblue, xshift=-0.3cm] at (1.8cm, -0.67cm) {IM};
\node[color=matplotlibblue, xshift=-0.3cm] at (1.8cm, -0.47cm) {IBF};
\node[color=matplotlibblue, xshift=-0.3cm] at (1.8cm, -0.87cm) {$\phi$};
\draw[line width=1mm] (0.5cm, -1cm) -- + (2cm, 0cm) node[anchor=north, pos=0.5] {Gun};


\foreach \x\l in {\Lone/{Cavity\\1}, \Ltwo/{Cavity\\2}, \Lthree/{Cavity\\3}, \Lfive/{Cavity\\4}} {
    \cavity{\x}{0cm}{\l}
}

\foreach \x\l in {\LSone/$\mathrm{ILS}_1$, \LStwo/$\mathrm{ILS}_2$, \LSthree/$\mathrm{ILS}_3$} {
    \solenoid{\x}
    \node[yshift=0cm, xshift=2\lslength-0.2cm, align=center, color=matplotlibblue] at (\x + \lslength/2, -0.87cm) {\l};
}

\draw[dotted] (2.8cm, -0.8cm) -- +(0, 1.6cm) node[yshift=0.25cm] {0m};


\draw[dotted] (12cm, -0.8cm) -- +(0, 1.6cm) node[yshift=0.25cm] {26m};

\begin{scope}[very thick]
    \draw (\gunwidth + 2.5cm, 0) -- (\Lone, 0);
    \draw ( {\Lone + \cavitylength - 0.045cm}, 0) -- (\Ltwo, 0);
    \draw (\Ltwo + \cavitylength - 0.045cm, 0) -- (\Lthree, 0);
    \draw (\Lthree + \cavitylength - 0.045cm, 0) -- (\Lfive, 0);
    \draw (\Lfive + \cavitylength - 0.045cm, 0) -- (11.1cm, 0) edge[dotted] (11.8cm, 0);
    \draw (11.8cm, 0) -- (12cm, 0);
\end{scope}
\draw[line width=1mm] (11.9cm, -1cm) -- + (0.1cm, 0cm) node[anchor=north, pos=0.5] {Experiment};

\end{scope}


\node[inner sep=0pt] (russell) at (0,3)
    {\includegraphics[width=12cm, height=7.8cm]{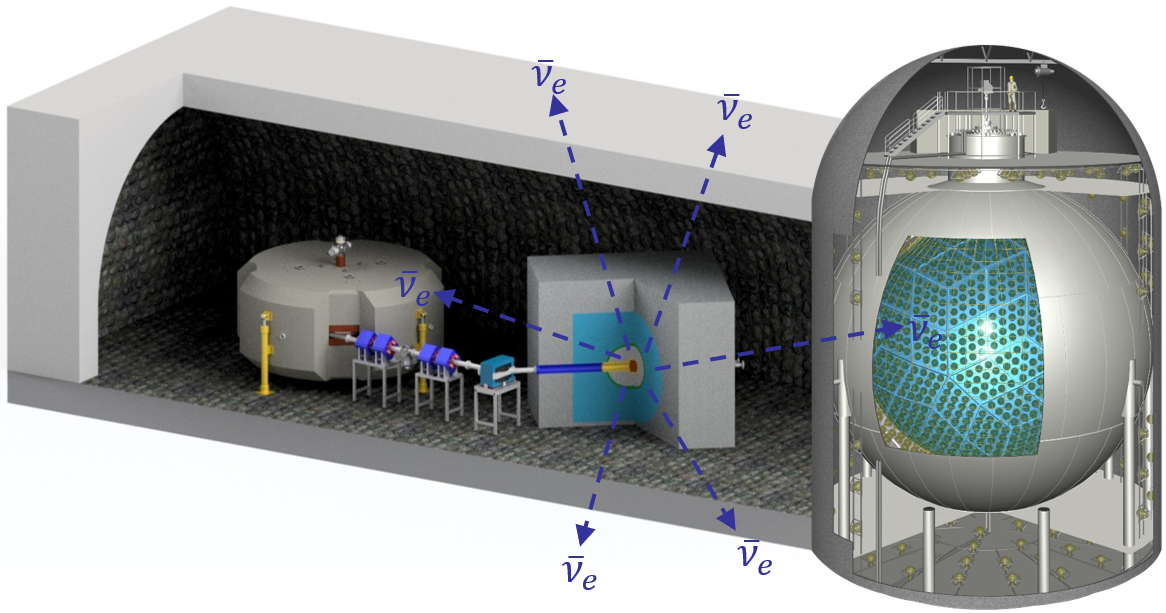}};

\node[inner sep=0pt] (russell) at (5,3)
    {\includegraphics[width=12cm, height=7.8cm]{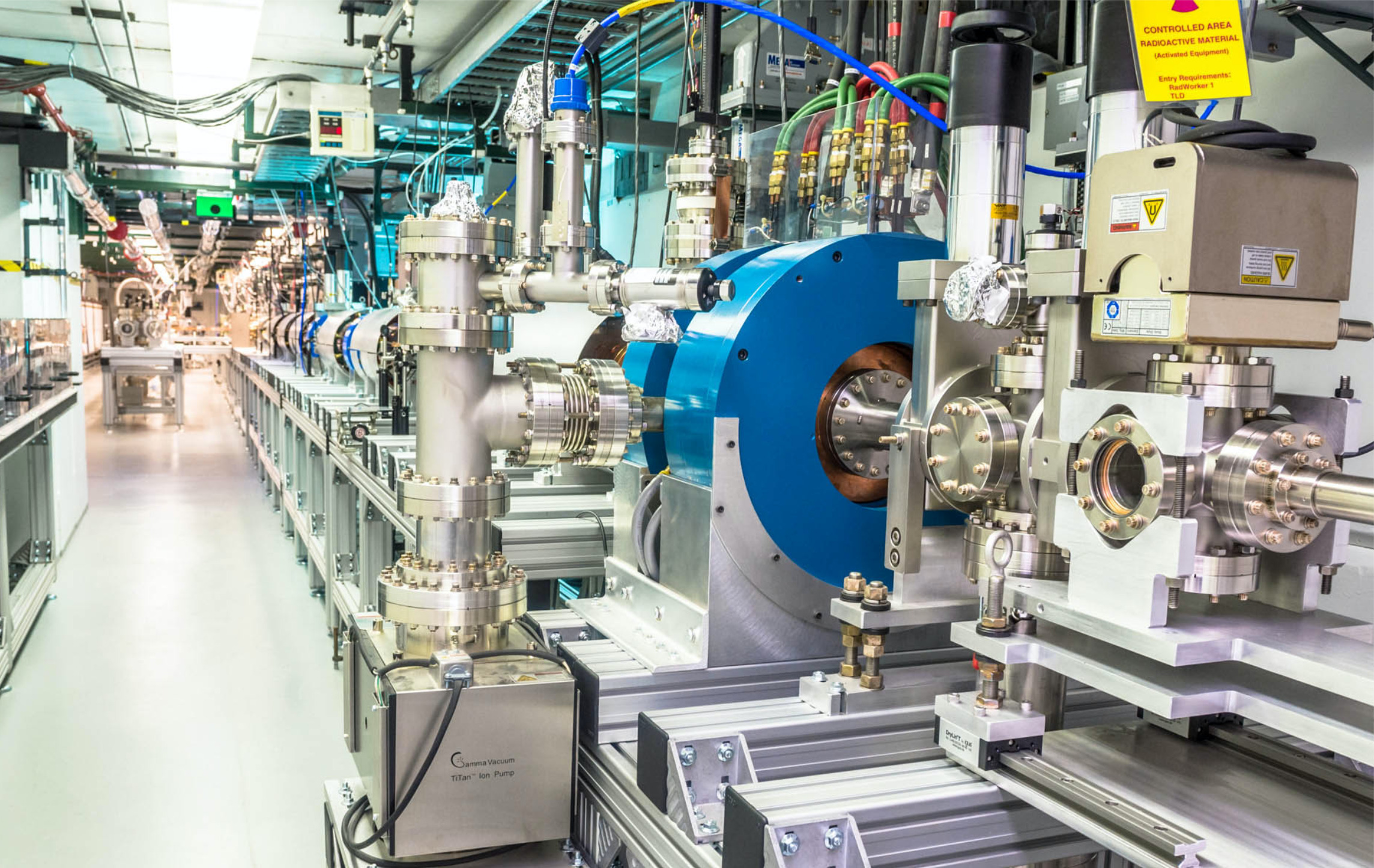}};

\node[inner sep=0pt] (russell) at (-1.8,4) {\Large\bf a};
\node[inner sep=0pt] (russell) at (-1.8,1.2) {\Large\bf b};
\node[inner sep=0pt] (russell) at (3.2,4) {\Large\bf c};
\node[inner sep=0pt] (russell) at (3.2,1.2) {\Large\bf d};
\end{tikzpicture}
}
\caption{Overview of the use-cases. \textbf{a} Artistic representation of the \isodar\ experiment at KamLAND (Kamioka Observatory, Japan), from left to right the cyclotron, target and detector are depicted. \textbf{b} A schematic of the \isodar\ cyclotron with relevant parameter. \textbf{c} The AWA facility at Argonne National Laboratory (US), pictured from the gun downstream. \textbf{d} Schematic of the relevant parts of the AWA machine.}
\label{fig:overview}
\end{figure}

Advances in accelerator science and technology have enabled discoveries in particle physics and other fields --- from chemistry and biology to medical applications --- for more than a century \cite{accel-phys-tod-2020}, with no end in sight \cite{no-front}.
Particle accelerators consist of a multitude of building blocks, and the relationship between changes in machine settings, known as \dvars, and the corresponding particle-beam response is often non-linear. For this reason, the development and operation of a particle accelerator rely heavily on computational models. These models use first principles of physics to state the equations of motion, while numerical algorithms are then employed to solve them. These models provide valuable insight, but their high computational cost is prohibitive for many applications. For example, a single simulation of the Argonne Wakefield Accelerator (AWA) (\figref{fig:overview}) takes approximately ten minutes with the high-fidelity physics model Object-Oriented Parallel Accelerator Library (\opal) \cite{opal}, which renders such models unsuitable for real-time usage. In order to achieve optimal operation points of these machines,  \gas\ (GAs) are typically the method of choice to solve multi-objective optimisation problems \cite{PhysRevAccelBeams.20.033401, PhysRevAccelBeams.22.054602, PhysRevAccelBeams.22.064602, PhysRevAccelBeams.22.122001}. However, an optimisation requires sometimes thousands of model evaluations. For this reason, the time-to-solution can easily be in the order of days. 
Recent work~\cite{ml_speedup} addressed this issue by training a data-driven surrogate model to approximate the computer model at a single position of interest. Evaluating such a surrogate model takes less than a second, and results in predictions that are very close to the the high-fidelity physics model. 

Related to these works, Scheinker et al.\ developed and demonstrated an online multi-timescale multi-objective optimisation algorithm that performs real-time feedback on particle accelerators \cite{doi:10.1063/5.0003423}. In contrast to our contribution they demonstrate a general control theoretical motivated feedback algorithm, capable to perform online multi-objective optimisation. Bayesian optimisation in combination  with particle-in-cell simulations \cite{PhysRevLett.126.104801} is used to tune a plasma accelerator autonomously to the beam energy spread to the sub-percent at a given energy and intensity. 

In this work we introduce two new surrogate models, a forward model and an invertible model. Both are designed to  model the machine at any position, overcoming a main limitation of existing approaches. Besides, the proposed methods to build those models are general enough to be applicable to any kind of particle accelerators. We demonstrate the generality of the Ansatz by considering a linear accelerator and a quasi circular machine i.e.\ a Cyclotron. Furthermore, the fast surrogate models enable optimisations on time scales suitable for on-line and real-time multi-objective optimisations of particle accelerators.

The forward model resembles existing models described e.\ g.\ in \cite{ml_speedup}. While previous work focused on approximating \opal\ only at one position (usually at the end) of the accelerator (as in \cite{ml_speedup}), our new model provides an approximation to \opal\ at a multitude of positions along the accelerator, allowing us to estimate beam properties along the entire accelerator depicted in \figref{fig:overview}\textbf{d}. To achieve this goal, our model takes both the \dvars\ and the position along the accelerator as input and predicts the beam properties at that position. Consequently our model is a complete replacement for \opal\ and does not limit our options of objectives and constraints for beam optimisations. Existing models are not capable of optimising the beam properties at multiple positions at the same time, but our forward model enables such sophisticated optimisations, which we demonstrate empirically.

The invertible model is capable of solving both the forward and the inverse problem by providing two kinds of prediction, called the forward and the inverse prediction. The forward prediction approximates \opal, i.\ e.\ it calculates the beam properties corresponding to a given machine configuration. The inverse prediction takes a user-imposed target beam and returns a machine configuration that realises an approximation to such a beam.
To incorporate expert intuition and knowledge in the optimisation, the result of the inverse prediction initialises a GA-based multi-objective optimisation, which then needs fewer generations to converge than when the usual random initialisation is used.

We demonstrate the capabilities of the two surrogate models with 
 the two real-world examples depicted in \figref{fig:overview}.\ The first example is the AWA, a linear accelerator. The second one is the cyclotron used for the proposed Isotope Decay-at-Rest experiment, a proposed very-high-intensity  electron-antineutrino source (from here on called the \isodar\ machine). We develop a \forwardmodel\ and an \invertiblemodel\ for both of these fundamentally different accelerators. Furthermore, we use them for a multi-objective accelerator optimisation to find machine configurations for desired beam properties. 

{\bf Physics Models and Datasets.}
For both accelerators we use the same underlying physics model, based on the \opal\ accelerator simulation framework. The dynamics of an accelerator are given by a function
\begin{align*}
    \myvec{f}: \mathbb{R}^m \times \mathbb{R} &\rightarrow \mathbb{R}^n\\
    \myvec{f}(\myvec{x}, s) &= \myvec{y}(s),
\end{align*}
where the quantity $\myvec{x}$ represents a vector of machine settings (\dvars), the scalar $s \in \mathbb{R}$ denotes the position along the accelerator, and the vector $\myvec{y}(s) \in \mathbb{R}^n$ denotes the beam properties (\qois). The function 
$\myvec{f}$ represents the \opal\ high-fidelity physics model.

Our surrogate models learn the relationship between \dvars\ and \qois\ based on examples, which means that we need to incorporate the physics of an accelerator into a dataset. We achieve this by sampling the space of \dvars\ randomly and evaluating the sampled configurations of interest at multiple positions along the machine. One sample in the dataset consists of the \dvars, a position along the machine, and the corresponding \qois. 

{\bf Specifics of the Argonne Wakefield Accelerator Model.}
The AWA accelerator \cite{AWAPower} consists mainly of an electron gun, several radio-frequency cavities for accelerating the electrons, and magnets for keeping the beam focused. The electron gun generates bunches that form a train of Gaussians in longitudinal direction, separated by a peak-to-peak distance $\lambda$. Other electron-gun variables are the gun phase $\phi$, the charge $Q$, the laser spot size (SIGXY), and the current in the buck focusing and matching solenoids (BFS and MS). 
For details we refer to \figref{fig:overview}{\bf{d}}.\
 In total, we have nine \dvars, denoted $\myvec{x} \in \mathbb{R}^9$, that define the various operation points of the AWA. The ranges of $\myvec{x}$ are determined by the physical limitations of the accelerator and are listed in \tabref{tab:awa_dvar}. The \qois\ (QoIs), $\myvec{y} \in \mathbb{R}
^{8}$, are: the transversal root-mean-square beam sizes $\sigma_x, \sigma_y$, the transversal normalised emittances $\epsilon_x, \epsilon_y$, the mean bunch energy $E$, the energy spread of the beam $\Delta E$, and the correlations between transversal position and momentum $\mathrm{Corr}(x, p_x), \mathrm{Corr}(y, p_y)$. A summary and more details about the used QoIs are given in \tabref{tab:predicted_quantities}. 

We build a labeled dataset of the AWA using \opal\ and the latin hypercube procedure \cite{McKay_1979} to sample the search space of $\myvec{x}$, which results in a training/validation set of $18,065$ points and a test set of $913$ points.

{\bf Specifics of the \isodar\ Cyclotron Model.}
The simulations are based on the latest \isodar\ \cite{alonso, Bungau, doi:10.1063/1.5127681} 
cyclotron design, with the nominal 
beam intensity of \SI{5}{mA} \htp beam (equivalent to \SI{10}{mA} of protons). Ongoing modeling efforts of \isodar\ consider the radial momenta $p_{r0}$, the injection radius $r_0$, the phase $\phi_{rf}$ of the radio frequency of the acceleration cavities, and the root mean square beam sizes $\sigma_{x}, \sigma_{y}, \sigma_{z}$ at injection. The physical ranges of the machine settings are given in \tabref{tab:isodar_dvar}. The surrogate models will predict the transversal and longitudinal root mean square beam sizes $\sigma_x, \sigma_y, \sigma_z$, the projected emittances $\epsilon_x, \epsilon_y, \epsilon_z$, the beam halo parameters $h_x, h_y, h_z$, the energy $E$, the energy spread of the beam $\Delta E$, and the particle loss $N_l$ (see \tabref{tab:predicted_quantities}). For this example we have six machine settings $\myvec{x} \in \mathbb{R}^6$ and twelve \qois\ $\myvec{y} \in \mathbb{R}^{12}$. All the samples are taken at turn 95, in the vicinity of the extraction channel. 

The labeled dataset, obtained with \opal,  consists of $5,000$ random machine configurations. Among these, $1,000$ samples are randomly selected and used as the test set, $800$ for the validation set, and the remaining $3,200$ samples form the training set. 

\section*{Results}
\begin{figure}
    \centering
    \includegraphics[width=\textwidth]{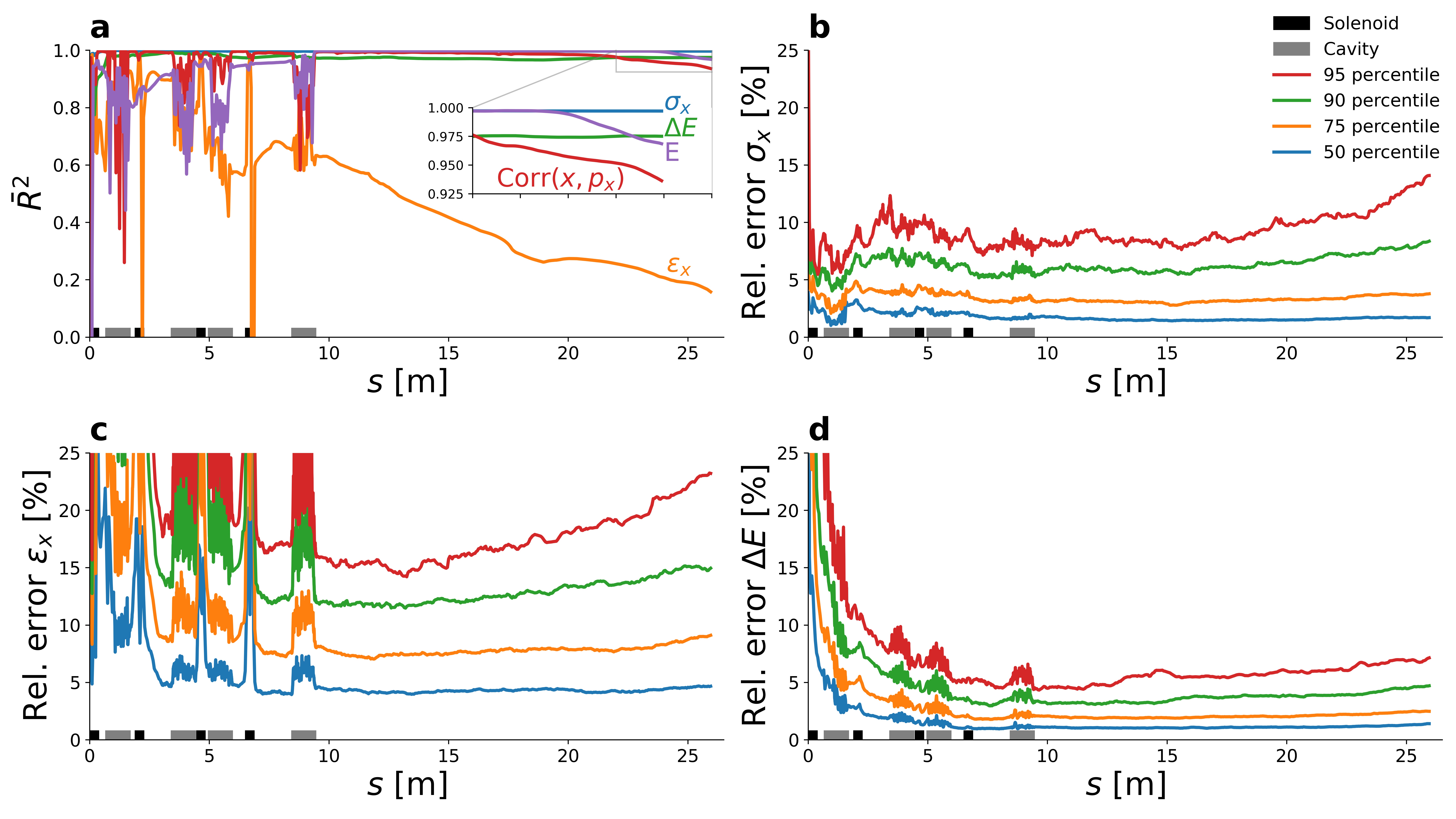}
    \caption{Prediction metrics for the AWA forward surrogate model. The plots show the adjusted coefficient of determination (\textbf{a}) and the relative prediction error (\textbf{b - d}) as functions of the longitudinal position. All values are calculated on the test set, i.\ e.\, on samples that have not been considered when choosing the parameters. We omit the curves for the QOIs in $y$ direction for the sake of readability because they are close to the ones in $x$ direction.}
    \label{fig:awa_forward_performance}
\end{figure}

{\bf Forward and Invertible Models.}
We assess the quality of the surrogate models by evaluating the adjusted coefficient of determination ($\bar{R}^2$) and the relative prediction error at $95\%$ confidence on the test set; i.e., data, that has not been used during the development of the models.
For the AWA model, the $\bar{R}^2$ values for the forward model are very close to one at nearly all positions along the machine, except for the solenoids and cavities. As a consequence, the variance of the dataset is explained well, as can be seen in \figref{fig:awa_forward_performance}{\bf{a}}. Only the variance of the emittance is not captured well, due to a numerical artifact arising from the 
choice of coordinate system in the solenoids and cavities respectively. However, the relative prediction error for this quantity is still less than $25\%$ for $95\%$ confidence, as depicted in \figref{fig:awa_forward_performance}{\bf{c}}. The relative errors for the other quantities are even lower than $15$ and $10\%$ for $\sigma_x$ (\figref{fig:awa_forward_performance}{\bf{b}}) and  $\Delta E$ (\figref{fig:awa_forward_performance}{\bf{d}}), respectively.

For the forward surrogate model of IsoDAR, the variance of the dataset is captured very well with a minimum $\bar{R}^2$ value of $0.95$ among all predicted values, except for the quantities in $z$-direction, here it ranges between $0.82$ and $0.86$. The relative prediction error of the forward model is at most $5.6\%$.

Since we are mainly interested in the beam parameters, we evaluated the invertible model by performing first the inversion, computing machine settings from beam parameters, with the invertible model and then a forward prediction with the forward model. For the AWA model the error at $95\%$ confidence is almost $40\%$ for the beam size, emittance, and energy spread. Nevertheless, the sampling error is only $20\%$ for three out of four target beams in the test set. The energy, on the other hand, is obtained almost with perfect accuracy.

For the IsoDAR model, the relative errors for $95\%$ confidence range from $0.08\%$ to $11\%$.
The corresponding $\bar{R}^2$ values lie between $0.79$ and $0.96$ except for the longitudinal quantities, for which we obtained values between $0.65$ and $0.77$.
The detailed values for the IsoDAR model can be found in \tabref{tab:isodar_r2}, \figref{fig:isodar_forward_percentage_error} and \figref{fig:isodar_sampling_error}.  

{\bf Multi-objective Optimisation.}
To find good operation points for the accelerators, we solve multi-objective optmisation problems \cite{PhysRevAccelBeams.20.033401,PhysRevAccelBeams.22.054602,ml_speedup,bazarov}. The standard approach for that is visualized in \figref{fig:optimisations}{\bf{a}}: A physics based model, like OPAL, together with a GA and random initialization is used. We investigate two other approaches depicted in  \figref{fig:optimisations}{\bf{b}}. First, we use a GA together with the forward model, instead of OPAL, and a random initialisation. In the second approach, still the GA and the forward model are employed, but in addition the invertible model provides us with a good initial guess for the optimisation. 

{\bf Performance Metrics Optimisation.} Characterising an approximation to the Pareto front, so the set of optimal solutions,  encompasses two main aspects. First, we strive to converge quickly towards non-dominated \qois. The range of the objectives in the non-dominated set is chosen to analyse the convergence behaviour. Second, we demand that the non-dominated \qois\ are diverse, i.\ e.\, well spread out over the entire approximated Pareto front. The convex hull volume ($V_{ch}$), the number of solutions of the non-dominated set, as well as the generational distance \cite{Veldhuizen99multiobjectiveevolutionary} and the hypervolume difference (i.e., the difference between the hypervolume of the optimal solution and the hypervolumes of the Pareto front of each generation, with a reference point chosen close to the Nadir point \cite{Fonseca_2006}) assess this quality aspect. Finally, plotting projections of the non-dominating set provides a holistic view to compare the effect of the initialisation strategies (random vs invertible model initialisation) on both the convergence and the diversity of the non-dominated set.

{\bf Multi-objective Optimisation for AWA.}
\label{sec:awa_optim}
The constrained multi-objective optimisation problem 
\begin{alignat*}{3}
&\text{min}   &            &\sigma_x(26~\mathrm{m})   &\\
&\text{min}   &            &\epsilon_x(26~\mathrm{m}) &\\
&\mathrm{min} &            &\Delta E(26~\mathrm{m})   &\\
&\text{s.\ t.} &\sigma_x(s) &\geq 2~\mathrm{mm},       &\quad s \in 0, 0.25, 0.5, ..., 10~\mathrm{m}\\
&             &\sigma_x(s) &\leq 5~\mathrm{mm},       &\quad s \in 0, 0.25, 0.5, ..., 10~\mathrm{m}\\
&             &\left| \mathrm{Corr}(x, p_x) \right| &\leq 0.1,         &\quad \mathrm{at~26~m},
\end{alignat*}
represents a generic multi-objective optimisation task, found in many of the past and future AWA experiments \cite{AWAPower}. The optimisation of beam parameters was required at $26~\mathrm{m}$ for a pilot experiment performed in 2020, regarding a new scheme for electron cooling. In addition, there were multiple constraints at various positions along the accelerator.
We solve the optimisation problem twice, once initialising the GA population randomly and once initialising the GA population with the \invertiblemodel\ outcome, where we utilized target values for the beam parameters as listed in Tab.~\ref{tab:awa_target}.
\begin{figure}
    \centering
    \includegraphics[width=\textwidth]{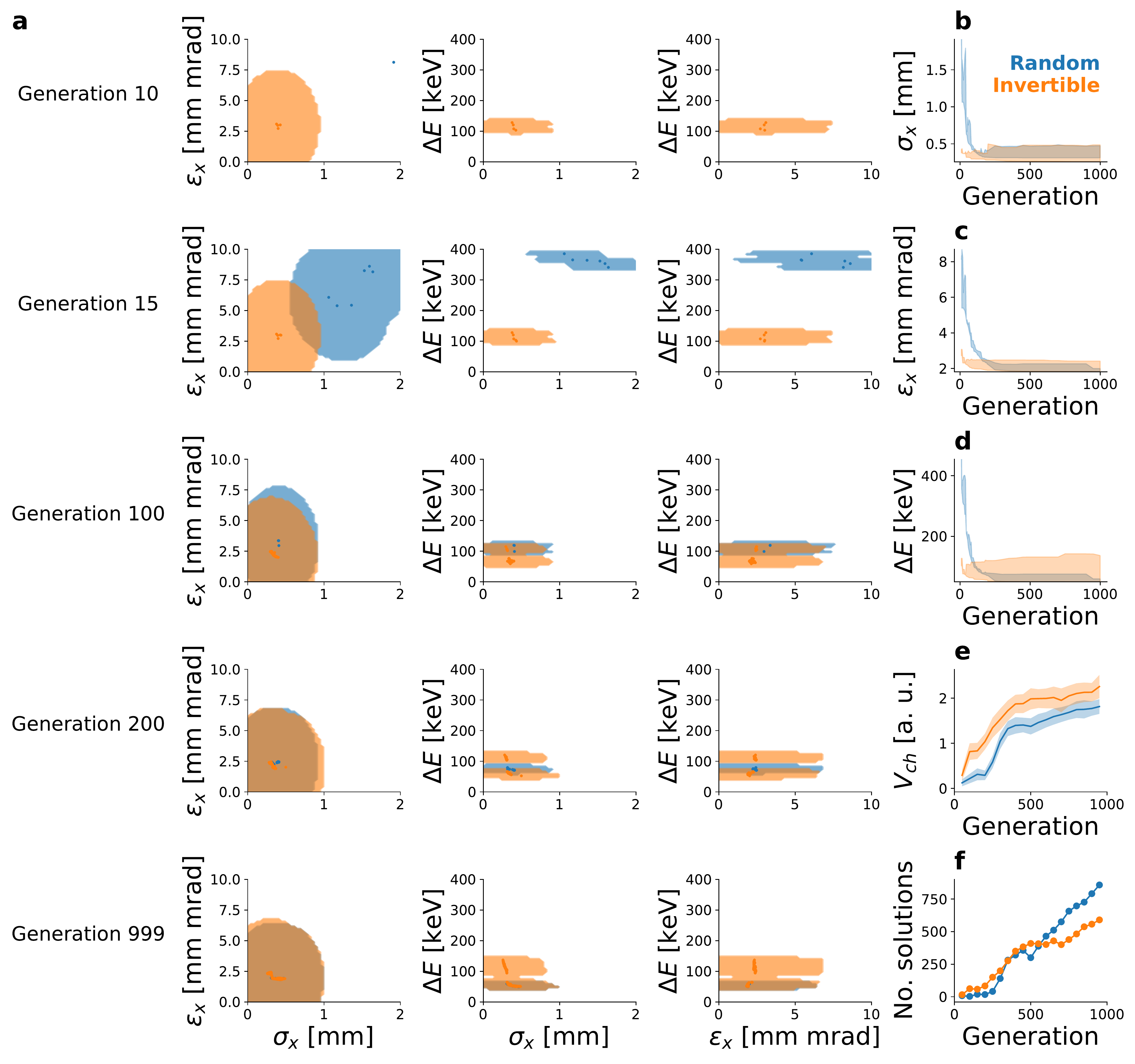}
    \caption{Convergence of the AWA optimisation. The figures show two-dimensional projections of the objective space (\textbf{a}), i.\ e.\ the \qois\ at \SI{26}{m}. The shaded area denotes the prediction uncertainty at $90\%$ confidence, the dots are the predictions themselves. Panels (\textbf{b - d}) depict the ranges of objective values among the non-dominated feasible solutions, the volume of the convex hull around the non-dominated solutions plus/minus Monte Carlo uncertainty (\textbf{e}), and the number of non-dominated solutions (\textbf{f}). The colour blue indicates that the optimiser is initialised randomly, orange indicates that the invertible model is used for the initialisation. All \qois\ are calculated using the forward surrogate model.}
    \label{fig:awa_convergence}
\end{figure}

The objective values for different generations are depicted in \figref{fig:awa_convergence}\textbf{a}. The optimisation initialised with the \invertiblemodel\  converged almost entirely after $10$ generations. If random initialisation is used, more than $100$ generations are needed to arrive at the same optimal configurations. The  ranges of the beam properties in the non-dominated set confirm this observation (\figref{fig:awa_convergence}\textbf{b-d}). Using the \invertiblemodel\ leads to a bigger hypervolume even when prediction uncertainty is accounted for (\figref{fig:awa_convergence}\textbf{e}). Furthermore, the optimisation initialised with the \invertiblemodel\ finds more solutions during the first $500$ generations (\figref{fig:awa_convergence}\textbf{f}). These two statements imply that the initialisation with the \invertiblemodel\ finds both more and more diverse non-dominated \dvars.

\begin{figure}
    \centering
    \includegraphics[width=\textwidth]{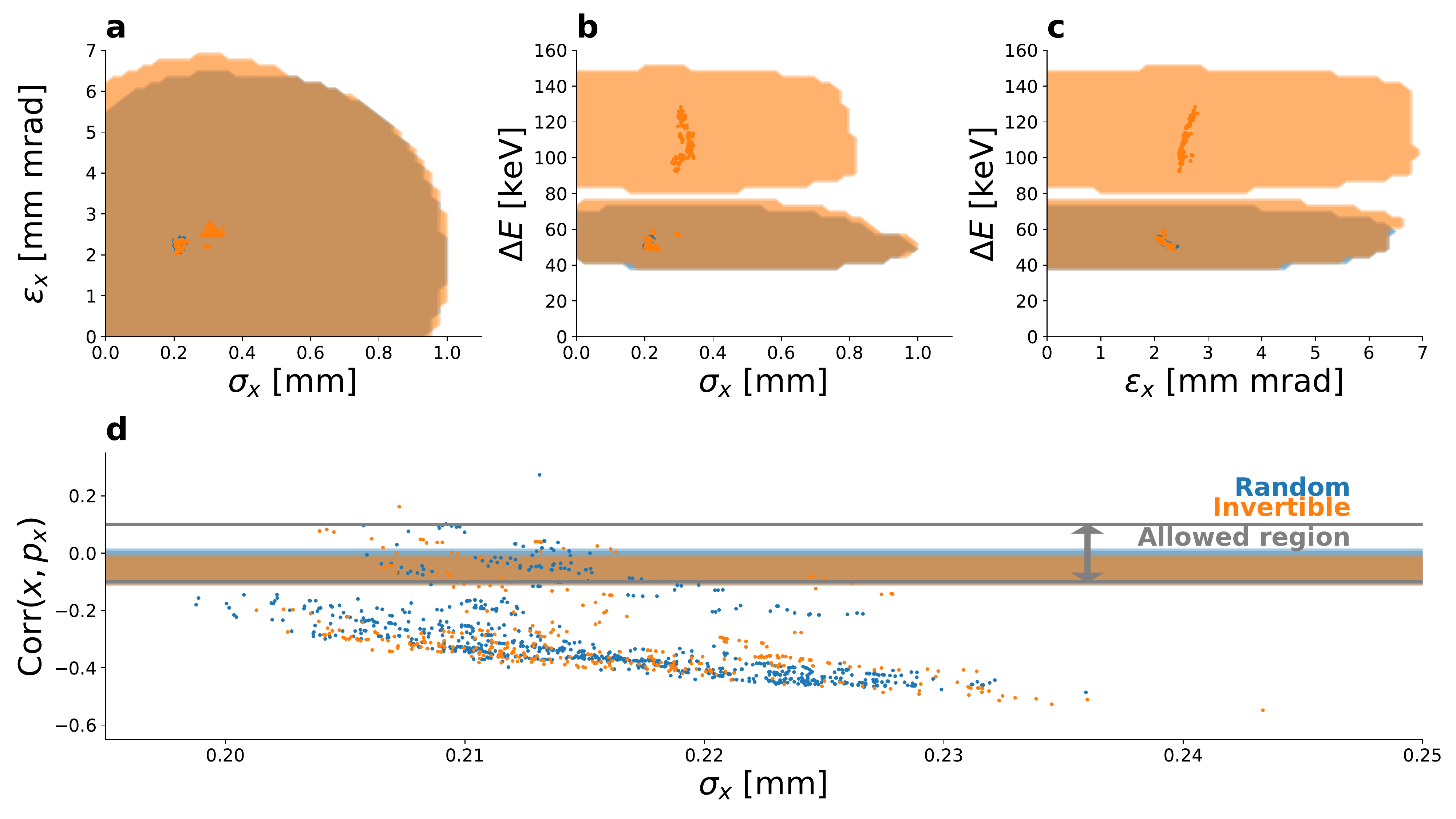}
    \caption{\opal\ validation of the non-dominated feasible machine settings among all $1000$ generations (dots) as calculated using the forward surrogate model. The colour blue marks optimisation with random initialisation, whereas orange indicates that the \invertiblemodel\ provided the initialisation. The dots are the results of an \opal\ run, the shaded areas mark values predicted by the forward model plus/minus uncertainty at $90\%$ confidence.}
    \label{fig:awa_validation}
\end{figure}

The non-dominated feasible machine settings are evaluated with \opal\ (dots in \figref{fig:awa_validation}\textbf{a}-\textbf{c}) to validate the predicted optimal objective values. All the \opal\ evaluated points lie within the region of uncertainty for the optimal objective values at $90\%$ confidence. 

In \figref{fig:awa_validation}\textbf{d} a deeper insight into the constraint on the correlation parameter ($Corr(x,P_x)$) is given. Although the surrogate models found many configurations lying outside the allowed region, both initialisation strategies lead to at least some feasible configurations. Importantly, no machine configuration in the training/validation set is feasible according to the optimisation problem. Nevertheless, some optimal points fulfill the correlation constraint. 


{\bf Multi-objective Optimisation for \isodar.}
For the \isodar\ case, we solve a two-objective optimisation problem, similar to that introduced by Edelen et al.~\cite{ml_speedup}. We aim to minimise simultaneously the projected  emittance $\epsilon_x$ and the energy spread $\Delta E$ of the beam without any constraints:
\begin{equation*}
    \begin{aligned}
        &\text{min} \quad & \Delta E\\
        &\text{min} \quad & \epsilon_x.\\
    \end{aligned}
\end{equation*}

As before, the optimisation problem is solved with a GA using the forward surrogate model. First a random initialisation is used and then an initialisation using the \invertiblemodel\ with the target beam parameters in \tabref{tab:isodar_target}. 


\begin{figure}[ht!]
    \centering
    \includegraphics[width=\textwidth]{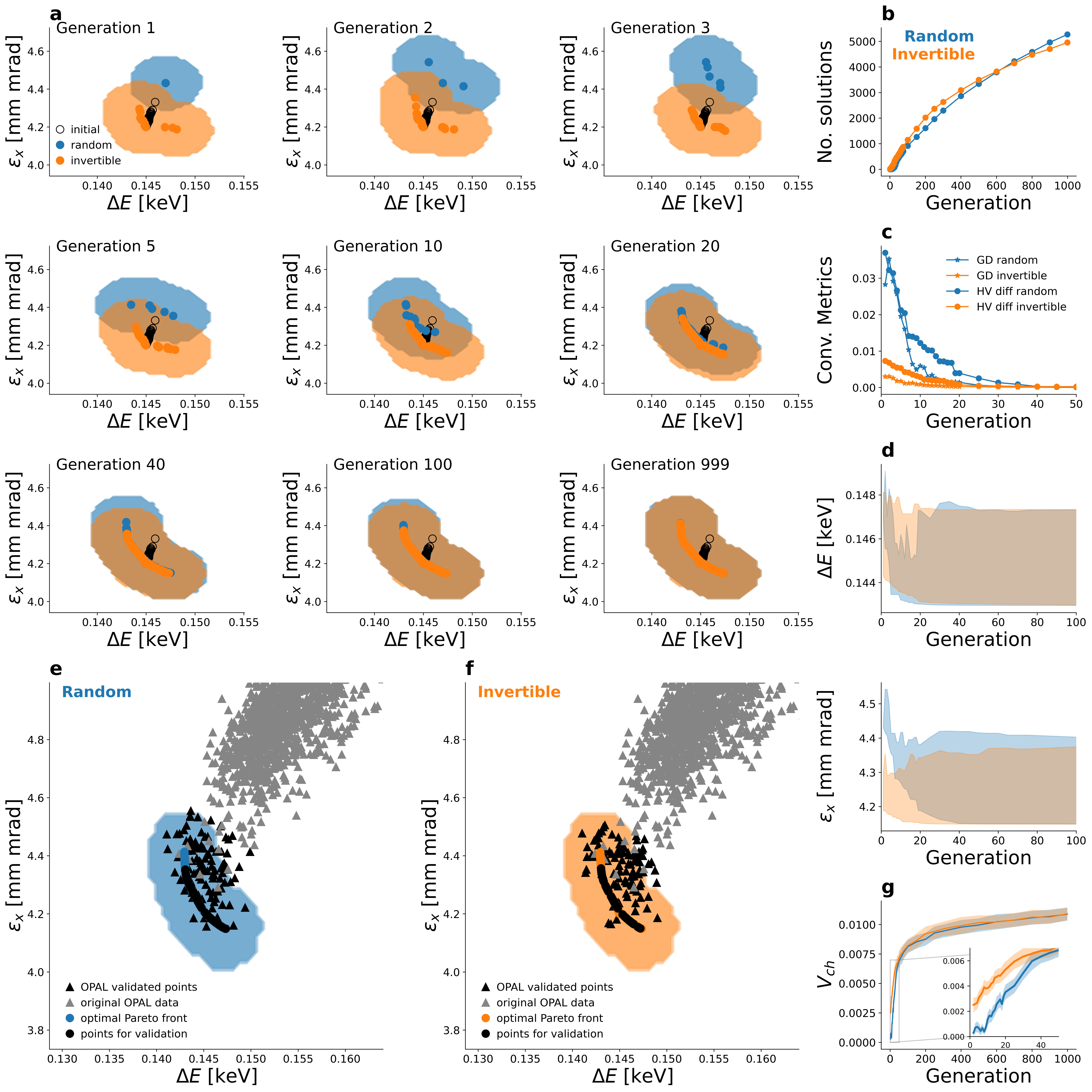}
    \caption{Convergence of the \isodar\ optimisation. The items coloured blue belong to the random initialisation, the orange ones indicate the usage of the \invertiblemodel\ for initialisation. \textbf{a} Two-dimensional projection of the objective space for different generations. The dots are the predictions and the shaded areas depict the prediction uncertainty at $90\%$ confidence as estimated on the test set. The black circles show the initial values for the biased optimisation. \textbf{b} Number of non-dominated solutions. \textbf{c} Convergence metrics: Generational Distance (GD) (dots) and Hypervolume Difference (HV diff) (stars) to compare the two optimisation approaches. \textbf{d} Range (minimum/maximum) of the non-dominated set in objective space.  \textbf{e},\textbf{f} Validation with \opal\: Both figures show the objective space with \opal\-validated points in black (triangles), the original \opal\ data points in grey (triangles), the 90 percentile (shaded area) of the optimal Pareto front (blue and orange dots) and 100 randomly chosen points of the optimal Pareto front for validation (black dots). \textbf{g} Estimated volume of the convex hull of the non-dominated set configurations in objective space using Monte Carlo uncertainty. The solid lines describe the mean values, the shaded areas mark the mean plus/minus one standard deviation.
}
    \label{fig:isodar_optimization}
\end{figure}

The convergence of the objective values for the two optimisation approaches is depicted in \figref{fig:isodar_optimization}\textbf{a}. Both approaches lead to the same non-dominated set. The optimisation with the \invertiblemodel\ initialisation converges faster and finds more optimal points already in the first few generations. The random initialisation approach needs approximately\ $40$ generations until it is at the same level. \figref{fig:isodar_optimization}\textbf{b} shows that both initialisation schemes lead in the end to nearly the same number of solutions.
The difference of the objective values for the two approaches in the first generations of the optimisation can also be seen in \figref{fig:isodar_optimization}\textbf{d}, which depicts the ranges from minimum to maximum values of the non-dominated set for the two objectives.

The performance-metrics plots \figref{fig:isodar_optimization}\textbf{c}, \textbf{g} show the behaviour of the two approaches quantitatively.
The generational distance for the \invertiblemodel\ initialisation starts at a much lower value. Therefore the non-dominated sets found in the first generation are already closer to the final non-dominated set than the ones found with the random initialisation. 
The lower hypervolume difference and the higher convex hull volume for the \invertiblemodel\ initialisation indicate that the initial non-dominated set is more widely spread, but these advantages vanish after approximately 40 generations. 



To validate the optimisation results, we randomly chose $100$ points of the non-dominated sets of both approaches and simulated them with \opal. We did not evaluate all non-dominated machine settings because of the high computational cost of \opal\ for the \isodar\ example.
As can be seen in \figref{fig:isodar_optimization}\textbf{e},\textbf{f}, nearly all of the \opal\ points lie within the 90-percent confidence region of the prediction with the test set, although only few points of the original \opal\ data set are in this area. 

{\bf Computational Advantages.}
We compare solving an optimisation problem with \opal\ to our approach of using a forward surrogate model instead of \opal. There are two quantities to be considered for the speedup analysis. First, the time-to-solution $t$ measured in hours and, second, the computational cost $c$ measured in CPU~hours. The detailed calculations for the improvement of both quantities can be found in \secref{sec:speedup} in the SI. Previous work \cite{adelmann-2020-1} also calculated speedups, but those calculations do not include the computational cost and the entire cost of developing a surrogate model because they neglect the hyperparameter scan. Moreover, the existing speedups refer to a model that is only capable of predicting the \qois\ at one position. Our calculations refer to our novel forward surrogate model, which is capable of predicting \qois\ at a plethora of positions, and do not only include the generation of the dataset, but also the training, hyperparameter tuning, and the cost of running the optimisations themselves. The \invertiblemodel\ reduces the number of generations. However, we calculate the speedup for a fixed number of generations, therefore the effect of the \invertiblemodel\ is not included in the speedup calculations.

For the AWA (see \tabref{tab:speedup_table}), we calculate the relative improvement ${t_\mathrm{\opal}}/{t_\mathrm{surr}} \approx 3.39$ in terms of time-to-solution and ${c_\mathrm{\opal}}/{c_\mathrm{surr}} \approx 1.83$ in terms of computational cost. This approach is more than three times \
faster and requires approximately $45\%$ less computational resources (see \eqnref{equ:cost_reduction} in the SI). For \isodar\ the benefits are even more pronounced: ${t_\mathrm{\opal}}/{t_\mathrm{surr}} \approx 640.00$ and ${c_\mathrm{\opal}}/{c_\mathrm{surr}} \approx 59.20$. Hence, the usage of a surrogate model saves more than $98\%$ of the computational cost and time resources compared to the approach using \opal. 


\section*{Methods}

\input{fig-optim}

{\bf Forward Model.}
\label{sec:forward_model}
The forward surrogate model $\tilde{f}$ (\figref{fig:optimisations}\textbf{c}) is a fast-to-evaluate approximation to the expensive OPAL-based physics model $f$
\begin{align*}
    \myvec{\tilde{f}}: \mathbb{R}^m \times \mathbb{R} &\rightarrow \mathbb{R}^n\\
    \myvec{\tilde{f}}(\myvec{x}, s) &= \myvec{\tilde{y}}(s)\\
    \mathrm{s.\ t.\ } \myvec{\tilde{f}}(\myvec{x}, s) &\approx \myvec{f}(\myvec{x}, s) \quad \forall \myvec{x}, s.
\end{align*}
While existing work \cite{ml_speedup} is restricted to approximating the accelerator at a single position $s$, usually at the end, we build a model for the AWA that approximates the dynamics of the machine at all positions. Without loss of generality, the \isodar\ model predicts the beam properties only at the end of the machine, as ongoing design work requires. We follow work by Edelen et al.\ \cite{ml_speedup} and use densely connected feedforward neural networks \cite{deeplearning} as candidates for the \forwardmodel. We focus only on architectures where all hidden layers are of the same width to simplify the hyperparameter scan. For the AWA model the hidden layers use the ReLU activation and the output layer uses the \texttt{tanh} activation function, whereas for the \isodar\ model \texttt{tanh} is used as an activation for the hidden layers and the output layer activation is linear. We use the mean absolute error (MAE) as the loss function for the AWA model and the mean squared error (MSE) for the \isodar\ model.

We optimise the trainable parameters using the Adam algorithm with default parameters. The non-trainable parameters are selected using grid search and can be found in \tabref{tab:parameters}, along with further information about the models. All the models (forward and invertible) are implemented using the \texttt{TensorFlow} framework \cite{tensorflow2015-whitepaper} in version \texttt{2.0}. For the hyperparameter scan of the IsoDAR model in addition the RAY TUNE library \cite{liaw2018tune} is used. The models are trained and evaluated on the Merlin6 cluster at the Paul Scherrer Institut, using 12 CPU cores for the AWA dataset and one CPU core for the \isodar\ case.


{\bf Invertible Model.}
 The \invertiblemodel\ (see \figref{fig:optimisations}\textbf{d}) is capable of performing two kinds of predictions: A forward prediction, approximating \opal, and an inverse prediction, aiming to solve the inverse problem: Given a target beam, the \invertiblemodel\ predicts a vector of \dvars\ such that the corresponding beam is a good approximation of the target beam.

We follow the ansatz by Ardizzone et al.~\cite{Ardizzone2018} and refer to their work for the details of the architecture of the inverse model. 
Ardizzone et al.\ build a neural network consisting of invertible layers, so-called affine coupling blocks (see \figref{fig:optimisations}\textbf{d}). Each of them contains two neural networks, sharing parameters. We decided that all hidden layers of the internal networks have the same number of neurons and that the internal networks of all affine coupling blocks have the same architecture. Each internal network consists of the same number of hidden layers of the same size. All hidden neurons of the internal networks use the ReLU activation function.

The solution of the inverse problem is not unique, hence a mechanism to select one solution is needed. This is realised by mapping the machine settings not only to the \dvars, but also to a latent space $\mathcal{Z} \subset \mathbb{R}^{d_z}$ of dimension $d_z$. This space follows a known probability distribution, which allows to sample random points in it. Sampling a point in the latent space corresponds to selecting one solution of the inverse problem. For this reason, the inverse prediction is also called sampling.

For technical reasons, an invertible neural network requires that the input and output vectors have the same length. This is generally not fulfilled.  If the input and/or output vectors are not big enough, vectors containing noise of small magnitude are added so that the total dimension of input and output vectors is $d$. We follow the advice of Ardizzone et al.~\cite{Ardizzone2018} and allow padding not only on the smaller vector, but on both, in order to increase the network width and therefore the model capacity. The total dimension $d$ is a hyperparameter. We denote the padding vectors $\myvec{x}_\mathrm{pad} \in \mathbb{R}^p, \myvec{y}_\mathrm{pad} \in \mathbb{R}^q$ and obtain
$d = m + 1 + p = n + 1 + \mathrm{dim}(\mathcal{Z}) + q.$

The mathematical formulation of the forward prediction is as follows:
\begin{align*}
    \myvec{\hat{f}}: \mathbb{R}^m \times \mathbb{R} \times \mathbb{R}^p &\rightarrow \mathbb{R}^n \times \mathbb{R} \times \mathcal{Z} \times \mathbb{R}^q\\
    \myvec{\hat{f}}(\myvec{x}, s, \myvec{x}_\mathrm{pad}) &= (\myvec{\hat{y}}, s, \myvec{z}, \myvec{y}_\mathrm{pad})\\
    \mathrm{s.\ t.\ } \myvec{\hat{y}} &\approx \myvec{f}(\myvec{x}, s).
\end{align*}
The inverse prediction is now written as
\begin{align*}
    \myvec{\hat{f}^{-1}}: \mathbb{R}^n \times \mathbb{R} \times \mathcal{Z} \times \mathbb{R}^q &\rightarrow \mathbb{R}^m \times \mathbb{R} \times \mathbb{R}^p\\
    \myvec{\hat{f}^{-1}}(\myvec{y}, s, \myvec{z}, \myvec{y}_\mathrm{pad}) &= (\myvec{\hat{x}}, s, \myvec{x}_\mathrm{pad})\\
    \mathrm{s.\ t.\ } \myvec{f}(\myvec{\hat{x}}, s) &\approx \myvec{y}.
\end{align*}

In order to improve the performance of the \invertiblemodel\, we use a best-of-$n$ strategy. For each inverse prediction (at inference time only), $n_\mathrm{tries}$ \dvar\ configurations are sampled, evaluated with the forward pass and the best configuration is chosen as the final prediction. Note that we do not need to rely on a \forwardmodel\, but only on the \invertiblemodel\ itself. We choose $n_\mathrm{tries} = 32$ for both the \isodar\ model and the AWA, as we observed no significant improvement for bigger values. All prediction errors are calculated with this particular choice.
The loss function consists of multiple parts: 
\begin{equation}
    \mathcal{L}_\mathrm{inv} = w_x \mathcal{L}_x + w_y \mathcal{L}_y + w_z \mathcal{L}_z + w_r \mathcal{L}_r + w_\mathrm{artificial} \mathcal{L}_\mathrm{artificial},
    \label{equ:inv_loss}
\end{equation}
 where the $\mathcal{L}_x$ loss ensures that the sampled machine setting vectors $\myvec{\hat{x}}$ follow the same distribution as the machine settings in the dataset, the $\mathcal{L}_z$ loss ensures that the latent space vectors follow the desired distribution. Both loss functions are realised as Mean Field Discrepancy \cite{gretton2012kernel}. The $\mathcal{L}_y$ loss is the mean squared error between $\hat{\myvec{y}}$ and $\myvec{y}_\mathrm{true}$, and the $\mathcal{L}_\mathrm{artificial}$ is the sum of the MSEs of both $\myvec{x}_\mathrm{pad}$ and $\myvec{y}_\mathrm{pad}$. The reconstruction loss $\mathcal{L}_r$ ensures that the inverse prediction is robust with respect to perturbations of small amplitude $\myvec{\epsilon}$
\begin{equation*}
    \mathcal{L}_r = \sum_{i = 1}^N \left\lVert \myvec{\hat{f}^{-1}} \left( \myvec{\hat{f}}(\myvec{x}_i, s_i) + \epsilon \right) - \myvec{x}_i \right\rVert^2.
\end{equation*}

The inverse prediction aims to generate machine settings $\hat{\myvec{x}} = \myvec{\hat{f}}^{-1}(\myvec{y}, s)_{1:m}$ such that the resulting beam $\myvec{\hat{y}} = \myvec{f}(\myvec{\hat{x}}, s)_{1:n}$ comes close to desired target-beam properties. To estimate the prediction accuracy of the models, we reproduce vectors $y$ from the test set. It is not reasonable to compare the sampled machine settings to the corresponding setting $\myvec{x}$ in the test set because multiple machine settings might realise similar beams. Instead, we perform a forward prediction with the forward surrogate model. This allows us to estimate to which beam properties the sampled configuration leads $\myvec{\tilde{y}} = \myvec{\tilde{f}}(\myvec{\hat{x}}, s)$. That quantity is used to describe the error of the inverse prediction. 

\section*{Discussion and Outlook}
We have introduced a novel flexible forward surrogate model that is capable of simulating high-fidelity physics models at a plethora of positions along the accelerator. 
We have shown that forward surrogate models reduce both the time (by a speedup factor of $3.39$ for the AWA and $640$ for \isodar) and the computational cost (by $45\%$ and $98\%$) of beam optimisations. The benefits increase if several optimisations need to be performed, because the time-consuming and computationally expensive part is the development of the model. Once the models are developed, the optimisations are almost free and can be performed in approximately $10~\mathrm{min}$ on $12$ cores. This allows experimenters to adapt to new situations by changing constraints and objectives and to rerun the optimisation with very little computational resources. Furthermore, the surrogate models themselves are useful in control-room settings, as they are fast enough to be used in a graphical user interface. This gives operators the possibility to quickly try out different machine settings. Despite of all these benefits of the surrogate model approach, one needs to be careful when applying them to problems with narrow constraints. Neural networks can only accurately model the machine-setting regimes represented in the data set on which they are trained. If the feasible region of an optimisation problem is narrow, the models will hardly be able to find configurations that fulfill all constraints. One solution to this issue is to sample the feasible region more densely instead of sampling the design space uniformly. This is difficult because usually the feasible region in design space is unknown a priori.
We have also presented an invertible model, which makes it possible to simulate the forward direction as well as the inversion, thus predicting machine settings that lead to desired beam parameters. The invertible model alone has the potential to support the design of accelerators, and it could also be used to implement a failure-prediction system or an adaptive control system. 
Another use-case of invertible surrogate models was demonstrated in this paper. Invertible surrogate models were shown to be able to bias the initialisation of a GA towards the optimal region, thereby incorporating prior knowledge and experience of experimenters. This reduces the number of generations needed for convergence of the GA when a forward surrogate model is used.

As the optimisation is almost free when a forward surrogate is employed, there is not much incentive to go through the labour of developing an \invertiblemodel. However, if \opal\ is used to evaluate all individuals, the reduced number of generations might reduce the time and cost of the optimisation significantly. Using \opal\ solves the problem of the underrepresented feasible space because \opal\ is capable of modelling the relevant physics of a particle accelerator. The approach of applying the biased initialisation to an \opal\ optimisation might combine both the advantage of needing fewer generations with the ability to accurately represent the feasible space. The accompanying savings might justify the cost of developing the \invertiblemodel. Further research is needed to investigate this prediction. Finally, the choice of the target vector is important, and additional research is needed to investigate its impact.

The societal impact of accelerator science and technology will continue, with no end in sight \cite{no-front,accel-phys-tod-2020,abs:isodar,alonso:isotopes}. With this research we add new computational tools and hence contribute to the quest of finding optimal accelerator designs and machine configurations, which in turn are likely to greatly reduce construction and operational costs and to improve physics performance. 

\printbibliography

\section*{Acknowledgements}
We acknowledge the help of Dr.\ John Power from AWA and
Dr.\ Daniel Winklehner from the \isodar\ collaboration.

\section*{Supplementary Material}
\appendix
\section{Predicted Quantities \& Model Fidelity}

{\bf Performance metrics.} We make use of the adjusted coefficient of determination for assessing the quality of the surrogate models, which is defined as
\begin{equation*}
    \bar{R}^2(s)~:=~1~-~\frac{\sum\limits_{i=1 \& s_i == s }^{N_\mathrm{test}}{} \left( \tilde{\myvec{f}}(\myvec{x}_i, s_i) - \myvec{y}_i \right)^2}{\sum\limits_{i=1 \& s_i == s}^{N_\mathrm{test}}{} \left( \myvec{y}_i - \bar{\myvec{y}} \right)^2} \cdot \frac{N_\mathrm{test} - 1}{N_\mathrm{test} - m - 1},
\end{equation*}
where $N_\mathrm{test}$ denotes the number of samples in the test set, $m$  is the number of \dvars\ $\myvec{x}_i \in \mathbb{R}^m$ and,  we only include samples where $s_i == s$.
This quantity can be interpreted as the fraction of variance in the data that is explained by the model. A perfect prediction corresponds to $\bar{R}^2 = 1$.

The prediction uncertainty at confidence $q$ is estimated by calculating the residuals of the prediction and take the absolute value over the test set samples. Then we calculate the $q$ percentile of these values. The uncertainty is calculated separately for each beam property $i$ and at every position $s$
\begin{equation*}
e_q = \left( q\percentile~\mathrm{at}~s \right)_i~\left| \tilde{y}_i - y_i \right|,
\end{equation*}
where the samples correspond to position $s$.

To measure the performance of the optimisation we use among others the convex hull volume. To compute that, the residuals over the test set are approximated with a Gaussian distribution. This allows us to sample new points around the predicted values and perform a Monte-Carlo estimate of the convex hull volume of the non-dominated set.


{\bf \isodar\ model fidelity.} 


 \begin{figure}[h!]
     \centering
     \includegraphics[width=\textwidth]{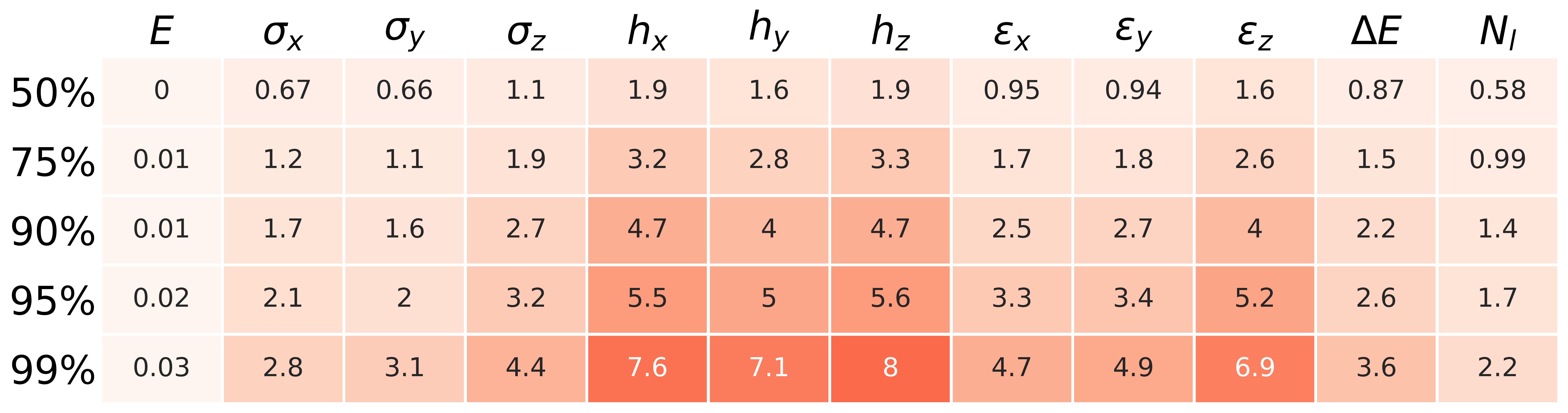}
     \caption{Percentiles of the relative prediction error for each quantity predicted by the forward \isodar\ model.}
     \label{fig:isodar_forward_percentage_error}
 \end{figure}
 \begin{table}[h!]
     \centering
     \resizebox{\textwidth}{!}{\begin{tabular}{lllllllllllll}
         \toprule
         &$E$ & $\sigma_x$ &  $\sigma_y$ &  $\sigma_z$ &  $h_x$ &  $h_y$ &  $h_z$ &  $\epsilon_x$ &  $\epsilon_y$ &  $\epsilon_z$  &     $\Delta E$ &  $N_l$ \\
 \midrule
         \forwardmodel &1.0 &  0.97 & 0.96 & 0.85 & 0.9 & 0.95 & 0.86 & 0.99 & 0.99 & 0.82 & 0.98&  1.0 \\
         \invertiblemodel FP &0.97 &  0.92 & 0.91 & 0.75 & 0.84 & 0.88 & 0.81& 0.93 & 0.95 & 0.69 & 0.94 & 0.9  \\
         \invertiblemodel IP& 0.96 &0.94 & 0.88 & 0.77 & 0.83 & 0.82 & 0.77& 0.91 & 0.91 & 0.65  &  0.91& 0.79  \\
         \bottomrule
     \end{tabular}}
     \caption{$\bar{R}^2$ values of the \isodar\ models. The values of the row "\invertiblemodel\ FP" refer to the forward prediction of the \invertiblemodel, whereas the values in "\invertiblemodel\ IP" refer to its inverse prediction. The latter ones are calculated with the help of the \forwardmodel.}
     \label{tab:isodar_r2}
 \end{table}

 \begin{figure}[h!]
     \centering
     \includegraphics[width=\textwidth]{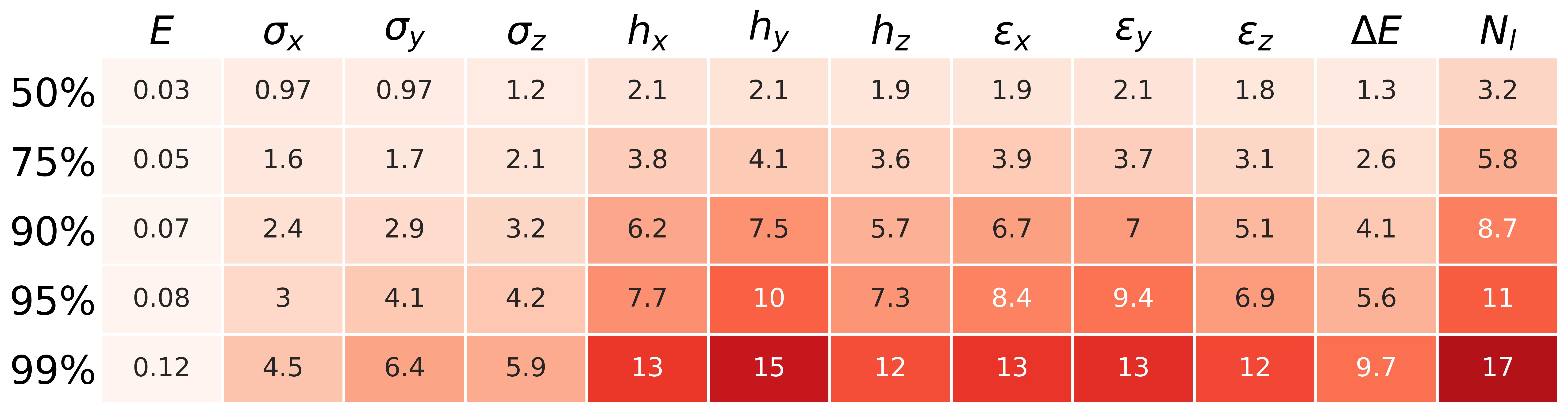}
     \caption{Prediction error for the inverse prediction of the \isodar\ model. This estimates how close the predicted machine settings get to the target beam. The rows represent the various confidence levels.}
     \label{fig:isodar_sampling_error}
 \end{figure}
\clearpage
{\bf Model summary.} The quantities predicted by our models are summarised in \tabref{tab:predicted_quantities}.

\begin{table}[h!]
    \centering
    \begin{tabular}{lcccc}
        \toprule
            & \multicolumn{2}{c}{AWA} & \multicolumn{2}{c}{\isodar}\\
            \cmidrule(lr){2-3} \cmidrule(lr){4-5}
            & Forward & Invertible & Forward & Invertible\\
        \midrule
        $E$ [MeV]                   & \checkmark & \checkmark & \checkmark & \checkmark \\
        $\Delta E$ [MeV]             & \checkmark & \checkmark & \checkmark & \checkmark \\
        $\sigma_x, \sigma_y$ [m]    & \checkmark & \checkmark & \checkmark & \checkmark \\
        $\epsilon_x, \epsilon_y$ []& \checkmark & \checkmark & \checkmark & \checkmark \\
        $\mathrm{Corr}(x, p_x), \mathrm{Corr}(y, p_y)$ [] & \checkmark & x & x & x \\
        $\sigma_z$  [m]             & x          & x          & \checkmark & \checkmark \\
        $\epsilon_z$  []           & x          & x          & \checkmark & \checkmark \\
        $N_l$  []                  & x          & x          & \checkmark & \checkmark \\
        $h_x, h_y, h_z$ []         & x          & x & \checkmark & \checkmark \\
    \end{tabular}
    \caption{Quantities predicted by the various models. Checkmarks \checkmark mean that the model predicts the quantity, while crosses x mean that the quantity is not predicted by the model.}
    \label{tab:predicted_quantities}
\end{table}

\pagebreak

\section{Parameter Ranges used in the Optimisation}
\label{Appendix:B}
\begin{table}[h!]
\resizebox{\textwidth}{!}{\begin{tabular}{lrrrrrrrrr}
Bound & IBF [A] & IM [A] & $\phi$ [\textdegree] & ILS$_1$ [A] & ILS$_2$ [A] & ILS$_3$ [A] & Q [nC] & $\lambda$ [ps] & SIGXY [mm]\\
\midrule
Lower & 450 & 100 & -50 &   0 &   0 &   0 & 0.3 & 0.3 &  1.5\\
Upper & 550 & 260 &  10 & 250 & 200 & 200 & 5   & 2   & 12.5\\
\bottomrule
\end{tabular}}
\caption{The \dvars\ and their ranges for the AWA models.}
\label{tab:awa_dvar}
\end{table}

\begin{table}[h!]
\center
\begin{tabular}{lrrrrrrrrrr}
Bound & $p_{r0}$ [$\beta \gamma$] & $r_0$ [mm] & $\phi_{rf}$ [\degree] & $\sigma_x$ [mm] & $\sigma_y$ [mm] & $\sigma_z$ [mm]\\
\midrule
Lower & 0.002254 & 115.9 & 283.0 & 0.95 & 2.85 & 4.75\\
Upper & 0.002346 & 119.9 & 287.0 & 1.05 & 3.15 & 5.25\\
\bottomrule
\end{tabular}
\caption{The \dvars\ and their ranges for the \isodar\ accelerator models.}
\label{tab:isodar_dvar}
\end{table}

From the $5,000$ samples for the \isodar\ model, $1,000$ were generated with machine configurations in the above ranges (\tabref{tab:isodar_dvar}). To increase the number of samples and improve the quality of the model, some ranges were narrowed. In detail $500$ samples were drawn with $\phi_{rf} \in [\ang{283.0}, \ang{284.0}]$ and another $3,500$ samples with additionally $r_0 \in [\SI{116.9}{\mm},\SI{119.9}{\mm}]$.


 \begin{table}[ht!]
     \centering
     \begin{tabular}{cccccc}
          \toprule
           $E$ & $\sigma_x$ & $\sigma_y$ & $\sigma_z$ & $h_x$ & $h_y$ \\
          \midrule
          $112.0$~MeV & $2.4$~mm & $2.1$~mm & $1.5$~mm & $4.5$ & $3.6$  \\
          \midrule[2pt]
           $h_z$ &$\epsilon_x$ & $\epsilon_y$ & $\epsilon_z$  & $\Delta E$&  $N_{l}$\\
          \midrule
          $3.4$ & $4.2$~mm~mrad & $4.5$~mm~mrad & $2.1$~mm~mrad  &  $146.2$~keV & $7090.0$\\
          \bottomrule
     \end{tabular}
     \caption{Target vector for the biased \isodar\ optimisation}
     \label{tab:isodar_target}
 \end{table}

 \begin{table}[ht!]
     \centering
     \begin{tabular}{ccccccc}
          \toprule
          E & $\sigma_x$ & $\sigma_y$ & $\epsilon_x$ & $\epsilon_y$ & $\Delta E$ & s\\
          \midrule
          $47.5$~MeV & $2$~mm & $2$~mm & 3~mm~mrad & 3~mm~mrad & 60~keV & 13.7~m\\
          \bottomrule
     \end{tabular}
     \caption{Target vector for the biased AWA optimisation.}
     \label{tab:awa_target}
 \end{table}
\pagebreak
\section{Model Parameters}
\begin{table}[h!]
    \centering
    \resizebox{\textwidth}{!}{\begin{tabular}{lllll}
        \toprule
                             & \multicolumn{2}{c}{AWA} & \multicolumn{2}{c}{\isodar}\\
                             \cmidrule(lr){2-3} \cmidrule(lr){4-5}
                             & Forward & Invertible & Forward & Invertible\\
         \midrule
         Modelled region                  & \makecell[tl]{$s \in [0, 26]~\mathrm{m}$ \\ (entire machine)} & $s \in [0, 13.7]~\mathrm{m}$ & EOM & EOM\\
         Preprocessing $\myvec{x}$        & Scale to $[-1, 1]$ & \makecell[tl]{\texttt{QuantileScaler} \\ Scale to $[-1, 1]$} & Scale to $[0, 1]$ & Scale to $[0, 1]$  \\ 
         Preprocessing $\myvec{y}$        & \makecell[tl]{Shift to be positive\\ Apply $\log()$\\ Scale to $[-1, 1]$}& \makecell[tl]{Clip $\epsilon_x \in [0, 200]~\mathrm{mm~mrad}$\\ Scale to $[-1, 1]$} & Scale to $[0, 1]$ & Scale to $[0, 1]$ \\
         Dimension of the latent space    & - & 1 & - & 1 \\
         Nominal Dimension                & - & 12 & - & 14 \\
         Distribution of the latent space & - & Unif(-1, 1) & - & Unif(-1, 1) \\
         Loss function                    & MAE & \makecell[tl]{$\mathcal{L}_\mathrm{inv}$ (\eqnref{equ:inv_loss})\\with weights:\\ $w_x = 400$\\ $w_y = 400$\\ $w_z = 400$\\ $w_r = 3$\\ $w_\mathrm{artificial} = 1$} & MSE & \makecell[tl]{$\mathcal{L}_\mathrm{inv}$ (\eqnref{equ:inv_loss})\\with weights:\\ $w_x = 400$\\ $w_y = 400$\\ $w_z = 400$\\ $w_r = 10$\\ $w_\mathrm{artificial} = 1$} \\
         Training algorithm               & Adam & Adam & Adam & Adam\\
         Learning rate                    & $10^{-4}$ & $10^{-4}$ & $10^{-4}$ & $10^{-3}$ \\
         Batch size                       & 256 & 256 & 256 & 8 \\
         Number of epochs                 & 56  & 15  & $5'000$ & 30 \\
         Architecture                     & $7 \times 500$ & $8 \times 3 \times 100$ & $6 \times 80$ & $5 \times 2 \times 60$ \\
         Activation of hidden neurons     & ReLU & ReLU & tanh & ReLU\\
         Number of trainable parameters   & $1'512'618$ & $688'192$ & $33'932$ & $91'340$ \\ 
         CPU cores for training           & $12$ & $12$ & $1$ & $1$\\
         Time for training                & $49$~h & $31$~h & $1$~h & $1$~h\\ 
         \bottomrule
    \end{tabular}}
    \caption{Parameters and properties of our surrogate models. EOM means end of the machine.}
    \label{tab:parameters}
\end{table}
Explanation of the network architectures: A feedforward network with $n_l$ hidden layers of width $n_w$ is denoted as an $n_l \times n_w$ network. An \invertiblemodel\ consisting of $n_b$ blocks, each containing internal networks of depth $n_d$ and width $n_w$, is noted as $n_b \times n_d \times n_w$.

The loss functions were described in the main text and are given here for completeness reasons: $\mathrm{MAE} = \frac{1}{N} \sum_{i=1}^N \left\lVert \myvec{\tilde{f}}(\myvec{x}_i, s_i) - \myvec{y}_i \right\rVert$, $\mathrm{MSE} = \frac{1}{N} \sum_{i=1}^N \left\lVert \myvec{\tilde{f}}(\myvec{x}_i, s_i) - \myvec{y}_i \right\rVert^2$, where $N$ denotes the number of samples over which the loss is calculated.
\clearpage

\section{Computational Details}
\label{sec:speedup}
\subsection*{Implementation}
The \forwardmodel\ and \invertiblemodel\ are implemented in \texttt{TensorFlow 2.0} \cite{tensorflow2015-whitepaper}. They are trained on 12 CPU cores for the AWA dataset and one core for the \isodar\ model. Training and evaluation were executed on the Merlin6 cluster at the Paul Scherrer Institute.
The hyperparameters of IsoDAR are found with a grid search using \cite{liaw2018tune}. 
\begin{algorithm}
\KwIn{Trained forward surrogate $\myvec{\tilde{f}}$.}
\KwIn{Number of generations $N_\mathrm{gen}$.}
\KwIn{Number of individuals in each generation $N_\mathrm{ind}$.}
\KwIn{Positions where to evaluate the objectives $\mathcal{S}_\mathrm{obj}$ and contraints $\mathcal{S}_\mathrm{con}$.}
\KwIn{NSGA-II function to advance by one generation $\mathrm{NSGA2}()$.}
\KwIn{Function that determines all non-dominated solutions from a set $\mathrm{non\_dominated}()$.}
\KwResult{Optimal set  of machine settings $X^*$.}
\KwResult{Optimal objectives as predicted by the forward surrogate $\tilde{Y}^*$.}
\KwResult{Optimal objectives as calculated by OPAL $Y^*$.}
\Begin{
	$X_0 \leftarrow \mathrm{initialise(N_\mathrm{ind})}$\\
	$X \leftarrow \{\}$\\
	$\tilde{Y} \leftarrow \{\}$\\
	\For{$i \in \{ 1, ..., N_\mathrm{gen} \}$}{
		Evaluate the current generation:\\
		$\tilde{Y}_i \leftarrow \{\}$\\
		\For{$\myvec{x} \in X_{i-1}$}{
			\For{$s \in \mathcal{S}_\mathrm{obj} \cup \mathcal{S}_\mathrm{con}$}{
				$\myvec{\tilde{y}}_{i, s} \leftarrow \myvec{\tilde{f}}(\myvec{x}, s)$\\
				$\tilde{Y}_i \leftarrow \tilde{Y}_i \cup \{\myvec{\tilde{y}}_{i, s}\}$\\
			}
		}
		$\tilde{Y} \leftarrow \tilde{Y} \cup \tilde{Y}_i$\\
		Build the new generation:\\
		$X_i \leftarrow \mathrm{NSGA2}(\tilde{Y}_i)$\\
		$X \leftarrow X \cup X_i$\\
	}
	Select the non-dominated individuals:\\
	$X^* \leftarrow \mathrm{non\_dominated}(X)$\\
	$\tilde{Y}^* \leftarrow \left\{ \myvec{\tilde{f}} (\myvec{x}, s) \middle| \myvec{x} \in X^*, s \in \mathcal{S}_\mathrm{obj} \right\}$\\
	Calculate the optimal beam parameters with OPAL:\\
	$Y^* \leftarrow \left\{ \myvec{f} (\myvec{x}, s) \middle| \myvec{x} \in X^*, s \in \mathcal{S}_\mathrm{obj} \right\}$
}
\caption{Optimisation using surrogate models.}
\label{alg:optimisation}
\end{algorithm}
We solve both the AWA and \isodar\ optimisation problems using the NSGA-II algorithm~\cite{nsga2} implemented by the Python library pymoo~\cite{pymoo}, with default parameters. The general algorithm is shown in Alg.~\ref{alg:optimisation}. The $\mathrm{NSGA2}()$ function is given the \qois\ for the current generation, calculates the values of the objectives and the constraints, and suggests the next generation of machine settings based on the results. When we refer to the optimal solutions of generation $g$, we mean the non-dominated feasible solutions that are found in the generations up to and including generation $g$. Notice that only the forward surrogate model is used to evaluate the objectives and constraints; \opal\ is used exclusively to train the models and validate the final optimal configurations.

We solve both optimisation problems in two ways: First, we initialise the first generation randomly by uniformly sampling machine settings from their ranges (see \tabref{tab:awa_dvar} and \tabref{tab:isodar_dvar}).\
 Second, the \invertiblemodel\ is used to bias the initialisation towards the optimal region. To achieve this, we provide a target vector of beam properties $\myvec{y}_t$ at position $s_t$ and ask the \invertiblemodel\ to sample the individuals in order to achieve this beam, see Alg.~\ref{alg:biased_init}. In other words, we guess what a good beam looks like at one position, and let the network calculate corresponding machine settings. This is what allows operators and experimenters to incorporate their experience and intuition into the optimisation.
 
\begin{algorithm}
\KwIn{Trained invertible surrogate $\myvec{\hat{f}}$.}
\KwIn{Number of individuals per generation $N_\mathrm{ind}$.}
\KwIn{Target vector $\myvec{y_t}$ at position $s_t$.}
\KwResult{First generation $X_0$.}
\Begin{
    $X_0 = \{\}$\\
    \For{$i \in \left\{ 1, ..., N_\mathrm{ind} \right\}$}{
        $\myvec{z} \leftarrow \mathrm{Unif}(-1, 1)$\\
        $\myvec{y}_\mathrm{pad} \leftarrow 5 \cdot 10^{-2} \cdot \mathrm{Normal}(0, 1)$\\
        $\myvec{x} \leftarrow \myvec{\hat{f}}^{-1} \left( \myvec{y}_t, s_t, \myvec{z}, \myvec{y}_\mathrm{pad} \right)$\\
        $X_0 \leftarrow X_0 \cup \{ \myvec{x} \}$
    }
}
\caption{Biased initialisation using the invertible surrogate model.}
\label{alg:biased_init}
\end{algorithm}



\subsection*{Speedup Calculation}
Let $t_o$ be the time needed for a single \opal\ simulation, running on $r_o$ cores. Let $t_t$ be the time to train a surrogate model on $r_t$ CPU cores, and $t_p$ the time needed to evaluate one individual of the optimisation (the p stands for prediction). Assume that the prediction takes place on the same number of cores as the training. Let $n$ be the number of unique machine configurations in the datasets. Since the development of a surrogate model involves a training, validation and test set, all of them are included in this number. Let $n_h$ be the number of hyperparameter configurations to be tried for the model. A single optimisation requires running $n_g$ generations, each consisting of $n_i$ individuals. A summary of the parameters along with their values for both models is depicted in \tabref{tab:speedup_table}. The runtime of \opal\ depends on the machine settings. The quantity is obtained by measuring the runtime for generating the training/validation set and then dividing by the number of unique machine settings in this set.

For all calculations, assume that we have infinite computational resources at our disposal. This means that all calculations can always exhaust the entire theoretically available parallelism. In practice, run times will usually be higher than the ones calculated using this assumption. The only step where the surrogate model is limited by computational resources is the generation of the dataset. This step is embarrassingly parallel, so the surrogate model can adapt perfectly to limited computational resources. If only \opal\ is used, however, the optimisation is also affected by limited resources. In the case where only one individual of a generation cannot be evaluated in parallel to the others, the time needed for the optimisation already doubles. This is because the next generation of the optimisation algorithm cannot start before the parent generation is fully evaluated\footnote{Assuming that the optimisation algorithm features no parallelism across generations, which is the case for the \texttt{pymoo} implementation.}. Therefore, our assumption of infinite parallelism favours the \opal-only approach. For this reason, our calculations lead to a lower bound for the speedup.

Now we can calculate the time-to-solution for an optimisation using the surrogate model. This task involves the creation of the dataset and the model training (including a hyperparameter scan). The former is embarrassingly parallel and we have unlimited computational resources, so all \dvar\ configurations are evaluated at once. The same argument applies for the hyperparameter scan. The time-to-solution for an optimisation with the surrogate model is given by
\begin{equation*}
    t_\mathrm{surr} = t_o + t_t + n_g \cdot t_p.
\end{equation*}
The associated computational cost is calculated as follows:
\begin{equation*}
    c_\mathrm{surr} = n \cdot t_o \cdot r_o + n_h \cdot t_t \cdot r_t + n_g \cdot n_i \cdot t_p \cdot r_t.
\end{equation*}
The time-to-solution for an \opal\ optimisation is calculated as follows:
\begin{equation*}
    t_\mathrm{\opal} = n_g \cdot t_o
\end{equation*}
The corresponding computation cost is calculated by
\begin{equation*}
    c_\mathrm{\opal} = n_g \cdot n_i \cdot t_o \cdot r_o.
\end{equation*}
Now we can determine the speedup in terms of execution time
\begin{equation*}
    \begin{aligned}
    \frac{t_\mathrm{\opal}}{t_\mathrm{surr}} &= \frac{n_g \cdot t_o}{t_o + t_t + n_g \cdot t_p}\\
      &= \frac{n_g}{1 + \frac{t_t}{t_o} + n_g \cdot \frac{t_p}{t_o}}\\
      &\approx \frac{n_g}{1 + \frac{t_t}{t_o}},
    \end{aligned}
\end{equation*}
where we have used $n_g \cdot \frac{t_p}{t_o} \ll 1$ for the approximation. This formula can be interpreted as follows: The bigger the runtime of \opal\ compared to the training time and the more generations we want to calculate for the optimisation, the higher the speedup will be.

The improvement in computational cost can be calculated by
\begin{equation*}
    \begin{aligned}
        C &= \frac{c_\mathrm{\opal}}{c_\mathrm{surr}}\\
          &= \frac{n_g \cdot n_i \cdot t_o \cdot r_o}{n \cdot t_o \cdot r_o + n_h \cdot t_t \cdot r_t + n_g \cdot n_i \cdot t_p \cdot r_t}
    \end{aligned}
\end{equation*}

The relative reduction in computational resources can then be calculated by
\begin{equation}
    \frac{c_\mathrm{surr} - c_\mathrm{\opal}}{c_\mathrm{\opal}} = \frac{c_\mathrm{surr}}{c_\mathrm{\opal}} - 1 = \frac{1}{C} - 1
    \label{equ:cost_reduction}
\end{equation}

Solving $n$ optimisation problems is equivalent to solving one problem using $n$ times the number of generations. This means that solving more optimisation problems is equivalent to using more generations. Asymptotically, the improvement in terms of computational cost scales like
\begin{equation*}
    C \xrightarrow{n_g \to \infty} \frac{t_o}{t_p} \cdot \frac{r_o}{r_t}.
\end{equation*}

\begin{figure}[!ht]
    \centering
    \includegraphics[width=\textwidth]{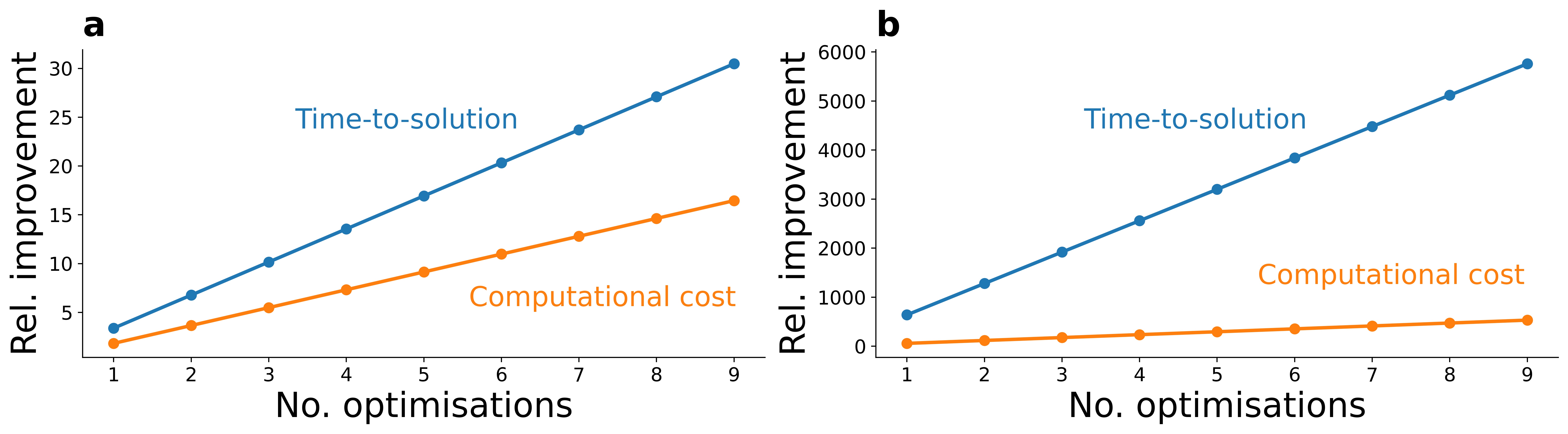}
    \caption{Relative improvement in terms of time-to-solution and computational cost for the AWA (\textbf{a}) and the \isodar\ (\textbf{b}) optimisations. The curves are calculated analytically based on measurements of the prediction time for the surrogate model and \opal. }
    \label{fig:speedup}
\end{figure}

The relationship of the relative improvement and the number of optimisations can be seen in \figref{fig:speedup}.

\begin{table}
    \centering
    \begin{tabular}{llll}
    \toprule
    Quantity & Symbol & AWA & \isodar \\
    \midrule
    Time for one \opal\ evaluation & $t_o$ & $\SI{10}{\minute}$ & $\SI{1.8}{\hour}$\\
    Time to train one \forwardmodel & $t_t$ & $\SI{49}{\hour}$ & $\SI{1}{\hour}$ 
    \\
    Time to predict one machine\footnote{For AWA, this means predicting the beam properties with a spacing of $5$~cm, i. e. at $520$~positions. For the \isodar\ accelerator, we model only $1$ position. The times are measured by measuring the prediction time for predicting $N_m$ machines and then dividing by $N_m$. For the AWA case, $N_m = 64$, for \isodar\ $N_m = 2^{20}$.} with the surrogate & $t_p$ & $\SI{21}{\milli \second}$ & $\SI{26}{\micro \second}$\\
    Number of unique machine settings in the dataset & $n$ & $21,000$ & $5,000$\\
    Number of hyperparameters to try & $n_h$ & 100 & $120$\\
    Number of CPU cores per \opal\ evaluation & $r_o$ & 4 & 1\\
    Number of CPU cores to train and evaluate the surrogate & $r_t$ & 12 & $1$\\
    Number of generations & $n_g$ & $1,000$ & $1,000$\\
    Number of individuals per generation & $n_i$ & $200$ & $300$\\
    \bottomrule
    \end{tabular}
    \caption{Parameters related to the speedup calculation.}
    \label{tab:speedup_table}
\end{table}

\end{document}